\documentclass[prb,twocolumn,longbibliography,superscriptaddress,preprintnumbers]{revtex4-2}

\usepackage[colorlinks,bookmarks=true,citecolor=blue,linkcolor=blue,urlcolor=blue] {hyperref}
\usepackage{dcolumn,graphicx,amsfonts,amsthm,bm,color,appendix,float}
\usepackage{braket}
\usepackage[normalem]{ulem}
\usepackage[version=4]{mhchem}
\usepackage{siunitx}
\usepackage{amsmath}
\usepackage{amssymb}
\usepackage{CJKutf8}
\usepackage{bbm}
\usepackage{subfigure}
\usepackage{color}
\usepackage{diagbox}

\theoremstyle{definition}
\usepackage{enumitem}

\usepackage[dvipsnames]{xcolor}
\definecolor{pal0}{rgb}{0.8941, 0.102 , 0.1098}
\definecolor{pal1}{rgb}{0.2157, 0.4941, 0.7216}
\definecolor{pal2}{rgb}{0.302 , 0.6863, 0.2902}
\definecolor{pal3}{rgb}{0.5961, 0.3059, 0.6392}
\definecolor{pal4}{rgb}{1.    , 0.498 , 0.    }

\begin{document}
\title{Topological magnons and domain walls in twisted bilayer MoTe$_2$}

\author{Wen-Xuan Qiu}
\affiliation{School of Physics and Technology, Wuhan University, Wuhan 430072, China}
\author{Fengcheng Wu}
\email{wufcheng@whu.edu.cn}
\affiliation{School of Physics and Technology, Wuhan University, Wuhan 430072, China}
\affiliation{Wuhan Institute of Quantum Technology, Wuhan 430206, China}

\begin{abstract}
We theoretically investigate the magnetic excitations in the quantum anomalous Hall insulator phase of twisted bilayer MoTe$_2$ at a hole filling factor of $\nu=1$, focusing on magnon and domain wall excitations. Using a generalized interacting Kane-Mele model, we obtain the quantum anomalous Hall insualtor ground state with spin polarization. The magnon spectrum is then computed via the Bethe-Salpeter equation, revealing two low-energy topological magnon bands with opposite Chern numbers. To further explore the magnon topology, we construct a tight-binding model for the magnon bands, which is analogous to the Haldane model. We also calculate the energy cost of domain walls that separate regions with opposite Chern numbers and bind chiral edge states. Finally, we propose an effective spin model that describes both magnon and domain wall excitations, incorporating Heisenberg spin interactions and Dzyaloshinskii-Moriya interactions. The coupling constants in this model are determined from the numerical results for magnons and domain walls. This model accounts for the Ising anisotropy of the system, captures the magnon topology, and allows for the estimation of the magnetic ordering temperature. Our findings provide a comprehensive analysis of magnetic excitations in twisted MoTe$_2$ and offer new insights into collective excitations in moiré systems.
\end{abstract}

\maketitle

\section{Introduction}
Quantum anomalous Hall insulators (QAHIs) are Chern insulators that do not require an external magnetic field, characterized by an insulating topological bulk state with a quantized Chern number and gapless chiral edge states~\cite{ColloquiumChang2023}. In transport measurements, QAHIs exhibit vanishing longitudinal conductance and quantized Hall conductance, akin to the quantum Hall effect but without the need for an applied magnetic field. QAHIs have been realized in thin films of magnetically doped topological insulators~\cite{CuiZuExperimental2013} and in the intrinsic magnetic topological insulator MnBi$_2$Te$_4$~\cite{YujunQuantum2020,LiuRobust2020}. Moir\'e superlattices provide a distinct platform to host QAHIs, where the moir\'e bilayers can be formed based on graphene systems~\cite{Cao2018,Cao2018a,Sharpe2019,Serlin2020,Chen2020,Polshyn2020,TschirhartImaging2021,Stepanov2021,Grovermosaic2022} or semiconducting transition metal dichalcogenides~\cite{LiQuantum2021,Xiaodong2023b,Yihang2023_integer,Park2023b,observation_Xu2023,Benjamin2024,RedekopDirect2024,AndersonTrion2024,JiLocal2024,ParkFerromagnetism2024,FerromagnetismPark2024,AnObservation2024}, spanning a broad range of material systems. 

In these moiré systems without any magnetic elements, QAHIs arise from the interplay between band topology and electron Coulomb interactions, which drive flatband ferromagnetism~\cite{Wu2019,Bultinck2020a,Bultinck2020,Xie2020,WuFengcheng2020Collective}. As a result, QAHIs spontaneously break time-reversal symmetry, supporting various low-energy collective excitations, such as excitons, magnons, domain walls, skyrmions, polarons, and trions~\cite{TopologicalSu2018,GuItinerant2019,KwanYves2021,GuItinerant2021,XieHongYi2023,FroeseTopological2024,WuFengcheng2020Quantum,EslamCharged2021,SkyrmionsKwan2022,KhalafBaby2022,KhalafSoft2020,WuFengcheng2020Collective,FerromagnetismAlavirad2020,TwistedBernevig2021,competitionKwan2021,TrionsSchindler2022,KondoHu2023,QuantumGeometry2024,MiguelDoping2024,TaigeDiverse2023}. Interesting properties have been revealed for these excitations. Exciton and magnons bands of QAHIs can be topological~\cite{TopologicalSu2018,GuItinerant2019,KwanYves2021,GuItinerant2021,XieHongYi2023,FroeseTopological2024}, and skyrmion excitations can bind an integer number of electrons~\cite{WuFengcheng2020Quantum, EslamCharged2021,SkyrmionsKwan2022,KhalafBaby2022}, reflecting the deep connection between the topological nature of QAHIs and their collective behavior.  
Theoretical studies have shown that these excitations are crucial for determining the stability, critical temperature,  low-energy charged quasiparticles, and optical responses of QAHIs~\cite{WuFengcheng2020Collective,WuFengcheng2020Quantum,EslamCharged2021,SkyrmionsKwan2022,KhalafBaby2022,QuantumGeometry2024}. In experiments, domain walls separating regions of QAHI with different Chern numbers or magnetization have been observed using local probes in graphene-based moiré materials~\cite{Grovermosaic2022,TschirhartImaging2021}. Additionally, signatures of charged spin skyrmions have been detected in magic-angle twisted bilayer graphene via scanning single-electron transistor measurements~\cite{YuSpin2023}.

In this work, we present a theoretical study of the magnetic excited states on top of the QAHI in twisted bilayer MoTe$_2$ ($t$MoTe$_2$) at a hole filling factor $\nu=1$. The $t$MoTe$_2$ system is particularly intriguing as it not only hosts QAHIs, but also their fractionalized counterpart, the fractional quantum anomalous Hall insulators (FQAHIs)~\cite{Xiaodong2023b,Yihang2023_integer,Park2023b,observation_Xu2023,Benjamin2024,RedekopDirect2024,AndersonTrion2024,JiLocal2024,ParkFerromagnetism2024,FerromagnetismPark2024,AnObservation2024}. While the ground-state properties of both QAHIs and FQAHIs in $t$MoTe$_2$ have been extensively studied~\cite{WangChong2024,fuliang_fractional,DongJunkai2023,BandAbouelkomsan2024,Electricallytuned2024,YuJiabin2024,LiuXiaoyu2024,InteractionDriven2023,IntertwinedSong2024,MagicMorales2024,ChengMaximally2024,MoirJia2024,PhaseSong2024,FanFengRen2024,TowardReddy2023,ZeroGoldman2023,PhaseWang2024,Majoranazero2024,VMbohao2024,AdiabaticShi2024}, our focus here is on the magnon and domain wall excitations in the QAHI phase at $\nu=1$, both of which are typical and important excitations in ferromagnetic systems.

Our study is based on an interacting Hamiltonian projected onto the first two bands in each valley of twisted bilayer MoTe$_2$ ($t$MoTe$_2$), where the single-particle part of the Hamiltonian is a generalized Kane-Mele model on a honeycomb lattice~\cite{Kane2005a,Kane2005,Wu2019,Hongyi2019_giant,Devakul2021,InteractionDriven2023}. Using a mean-field Hartree-Fock (HF) approximation, we theoretically obtain the QAHI phase with valley (equivalent to spin) polarization at $\nu=1$. We then calculate the momentum-dependent magnon spectrum, which can be interpreted as excitonic states involving spin  flips. The magnon spectrum and wavefunction are determined by solving the Bethe-Salpeter equation. The magnon spectrum is positive-definite, indicating the robustness and Ising ferromagnetic character of the QAHI phase in $t$MoTe$_2$. A key finding is that the first two magnon bands are topological, with opposite Chern numbers $(\pm 1)$. To further understand the magnon topology, we construct a tight-binding model for these two magnon bands, derived from two magnon Wannier states localized on the honeycomb lattice. This model takes a form similar to the Haldane model, providing a real-space perspective on the topological nature of the magnons. In addition, we perform a mean-field calculation for domain walls that separate regions of QAHIs with opposite spin polarization and Chern numbers. The chiral edge states bound to the domain walls are revealed, and the energy cost of the domain wall excitations is computed. Finally, we propose an effective spin model in the form of a nonlinear sigma model on the honeycomb lattice to provide a unified description of the magnon and domain wall excitations. The effective spin model has the in-plane and out-of-plane Heisenberg spin coupling, as well as the Dzyaloshinskii-Moriya interaction (DMI) terms between next-nearest neighbors on the honeycomb lattice. Here the DMI terms accounts for the topological gap opening between the first and second magnon bands. The values of the coupling constants in the effective spin model are determined from the numerical results for magnons and domain walls. We estimate the magnetic ordering temperature of the effective spin model and compare it to available experimental data for $t$MoTe$_2$.  Our work provides a systematic study of magnetic excitations in $t$MoTe$_2$ and offers new insights into collective excitations in moiré systems.

The paper is organized as follows. In Sec.~\ref{sec2}, we present the moir\'e Hamiltonian and band-projected interacting model, by which the mean-field QAHIs ground state is obtained. In Sec.~\ref{sec3}, we present the magnon spectrum calculated with the Bethe-Salpeter equation, after which a lattice model consisting of two magnon Wannier states is constructed. The numerical convergence of magnon energies is presented in Appendix \ref{appa}. The Berry curvature and Chern numbers of magnon bands are calculated, with technical details presented in Appendix \ref{appb}. In Sec.~\ref{sec4}, we calculate the mean-field band structure and energy for a superstructure with domain walls. In Sec.~\ref{sec5}, an effective spin model is constructed, which is used to characterize the above two excitations and estimate the magnetic ordering temperature. A microscopic justification of the effective spin model is given in Appendix \ref{appc}. We conclude with a summary and discussion in Sec.~\ref{sec6}.


\section{QAHI ground state}\label{sec2}

\subsection{Single-particle Hamiltonian}
We start by describing the single-particle moir\'e Hamiltonian of $t$MoTe$_2$ for valence band states in $\pm K$ valleys~\cite{Wu2019},
\begin{equation}\label{ctnh0_main}
\begin{aligned}
&\hat{\mathcal{H}}^{\tau}_0  =\left(\begin{array}{cc}
-\frac{\hbar^2\left(\hat{\boldsymbol{k}}-\tau\boldsymbol{\bm \kappa}_{+}\right)^2}{2 m^*}+\Delta_{+}(\boldsymbol{r}) & \Delta_{\mathrm{T,\tau}}(\boldsymbol{r}) \\
\Delta_{\mathrm{T,\tau}}^{\dagger}(\boldsymbol{r}) & -\frac{\hbar^2\left(\hat{\boldsymbol{k}}-\tau\bm\kappa_{-}\right)^2}{2 m^*}+\Delta_{-}(\boldsymbol{r})
\end{array}\right), \\
&\Delta_{\pm}(\boldsymbol{r}) = 2 V \sum_{j=1,3,5} \cos \left(\boldsymbol{g}_j \cdot \boldsymbol{r} \pm \psi\right), \\
&\Delta_{\mathrm{T,\tau}}(\boldsymbol{r})=w \left(1+e^{-i \tau\bm g_2 \cdot \boldsymbol{r}}+e^{-i \tau \bm g_3 \cdot \boldsymbol{r}}\right),
\end{aligned}
\end{equation}
where the $2\times2$ Hamiltonian $\hat{\mathcal{H}}^{\tau}_0$ is expressed in the layer-pseudospin space of the bottom($b$) and top($t$) layers. The index $\tau=\pm$ labels $\pm K$ valleys, which are also locked to out-of-plane spin $\uparrow$ and $\downarrow$, respectively. This is due to the spin-valley locking in MoTe$_2$~\cite{xiao2012}; therefore, spin and valley are used interchangeably in this work.  
$\Delta_\pm (\bm{r})$ is the layer-dependent potential with an amplitude $V$ and phase parameters $\pm \psi$, and $\Delta_{\mathrm{T,\tau}}(\boldsymbol{r})$ is the interlayer tunneling with a strength $w$. $\bm{r}$ and  $\hat{\bm{k}}$ are respectively, the position and momentum operators. $m^*$ is the effective mass. $\bm{\kappa}_{\pm}=\left[4\pi /(3 a_M)\right](-\sqrt{3}/2, \mp 1/2 )$ are located at corners of the moir\'e Brillouin zone (mBZ), and $\bm{g}_j=\left[4\pi /(\sqrt{3} a_M)\right]\{\cos[(j-1)\pi/3], \sin[(j-1)\pi/3]\}$ for $j=1,...,6$ are the moir\'e reciprocal lattice vectors, where $a_M\approx a_0/\theta$ is the moir\'e period, $a_0$ is the monolayer lattice constant, and $\theta$ is the twist angle.
Our study focuses on the physics of $\pm K$ valleys. In $t$MoTe$_2$, the $\Gamma$ valley valence states are deeper in energy and therefore lie away from the Fermi energy \cite{angeli2021gamma}, making them irrelevant to the low-energy physics we investigate.

The Hamiltonian $\hat{\mathcal{H}}^{\tau}_0$ respects the point group symmetry of $t$MoTe$_2$, which includes a threefold rotation $\hat{C}_{3z}$ around the out-of-plane $\hat{z}$ axis and a twofold rotation $\hat{C}_{2y}$ around the in-plane $\hat{y}$ axis that exchanges layers. Moreover, the low-energy Hamiltonian $\hat{\mathcal{H}}^{\tau}_0$ that only includes the lowest harmonic terms has an effective intravalley inversion symmetry~\cite{MoirJia2024}. Therefore,  the point group symmetry of $\hat{\mathcal{H}}^{\tau}_0$ is enlarged to $\hat{D}_{3d}$. In addition, $\hat{\mathcal{H}}^{\tau}_0$ preserves the time-reversal symmetry $\hat{\mathcal{T}}$ and valley $U(1)$ symmetry.

\begin{figure}[t]
\centering
\includegraphics[width=0.42\textwidth,trim=0 0 0 0,clip]{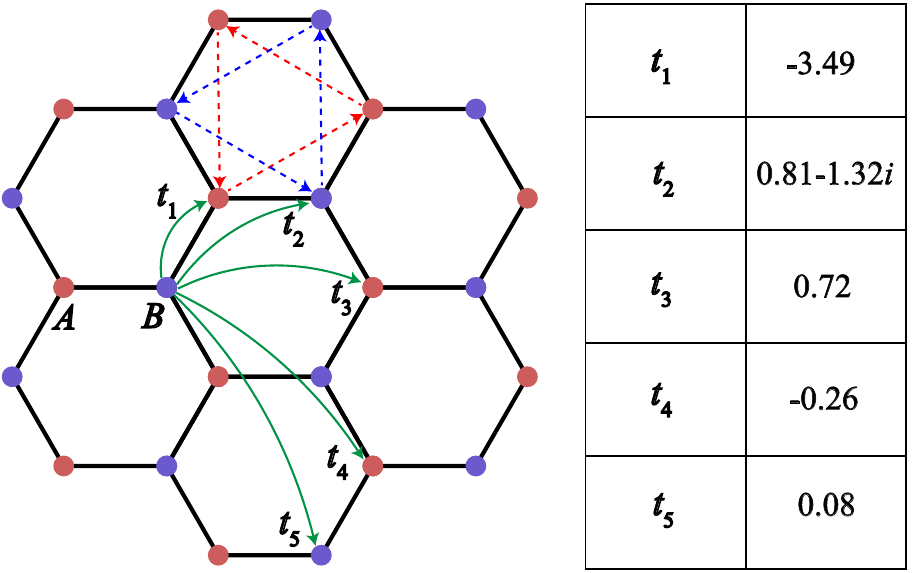}
\caption{Left panel: Schematic illustration of Kane-Mele model on honeycomb lattice. $t_n$ is the hopping parameter.  The next-nearest neighbor hoppings have a sublattice, spin, and direction-dependent phase factors, as illustrated by the dashed arrows. Right panel: Numerical values of $t_n$ in unit of meV at $\theta=3.5^\circ$, which are determined by constructing electronic Wannier states~\cite{Devakul2021,InteractionDriven2023} for the first two moir\'e bands of the continuum model in Eq.~\eqref{ctnh0_main}.}
\label{fig1}
\end{figure}

We take the following model parameters, $a_0 = 3.52 $\AA, $m^* = 0.6 m_e$, $V = 20.8$ meV, $\psi = -107.7^\circ$, $w = -23.8$ meV, as used in Refs.~\cite{WangChong2024,LiuXiaoyu2024,FanFengRen2024}, where $m_e$ is the electron bare mass. Under these parameters, the first two moir\'e valence bands of $\hat{\mathcal{H}}_{0}^{\tau}$ for $\theta$ around $3.5^{\circ}$ have opposite Chern numbers of $\pm 1$ in each valley, which can be represented in terms of Wannier orbitals at $A$ and $B$ sublattices on a honeycomb lattice~\cite{Wu2019,Devakul2021}, as shown in Fig.~\ref{fig1}. The Wannier states at $A(B)$ sites are, respectively, polarized to $t$ ($b$) layers.
Since we study holes doped into the valence bands, we work in the hole basis.
The tight-binding Hamiltonian based on the Wannier states in the hole basis can be expressed as,
\begin{equation}\label{thhtn}
\begin{split}
\hat{\mathcal{H}}_{\mathrm{KM}}
=\sum_\tau\sum_{\alpha\beta} \sum_{\bm R \bm R'}
t^\tau_{\alpha\beta}(\bm R'-\bm R) b^\dagger_{\bm R \alpha\tau} b_{\bm R'\beta\tau},
\end{split}
\end{equation}
where $b^\dagger_{\bm R \alpha\tau}$ ($b_{\bm R'\beta\tau}$) is the hole creation (annihilation) operator for the Wannier orbital at sublattice $\alpha$ ($\beta$) in the moir\'e unit cell $\bm R$ ($\bm R'$) and valley $\tau$. 
The form of $\hat{\mathcal{H}}_{\mathrm{KM}}$ is constrained by the $\hat{D}_{3d}$ point group symmetry, $\hat{\mathcal{T}}$ symmetry, and valley $U(1)$ symmetry.
The hopping terms and their numerical values are presented in Fig.~\ref{fig1} up to the fifth order, where $t_n$ represents the $n$th nearest-neighbor terms. We choose a gauge such that the nearest-neighbor hopping $t_1$ is real.  An important feature of the tight-binding model is that the second nearest-neighbor hopping terms are complex with spin, sublattice and direction-dependent phase factors $e^{i\phi_t \tau\nu_{ij}}$, where $\phi_t$ is a phase and  $\nu_{ij}=+1 (-1)$ if the hopping from the site $i$ to $j$ is along (against) the dashed arrows in Fig.~\ref{fig1}. The spin (valley) dependence of the phase factors breaks the spin SU(2) symmetry down to U(1) symmetry.
This complex hopping pattern is reminiscent of that in the Kane-Mele model (two copies of the time-reversal partner Haldane model) and plays a crucial role in the formation of Chern bands. Therefore, 
Eq.~\eqref{thhtn} can be understood as a generalized  Kane-Mele model on a honeycomb lattice with hopping beyond second nearest neighbors~\cite{Kane2005a,Kane2005}. 

By Fourier transformation and diagonalization, Eq.~\eqref{thhtn} can be written in the moir\'e band basis as,
\begin{equation}\label{H0mor}
\begin{aligned}
\hat{\mathcal{H}}_{\mathrm{KM}}&=\sum_{\boldsymbol{k}, \tau} \sum_{n=1,2} \mathcal{E}_{\bm k}^{n\tau} b_{\boldsymbol{k} n \tau}^{\dagger} b_{\boldsymbol{k} n \tau},
\end{aligned}
\end{equation}
where $b_{\bm k n \tau}^{\dagger}$ ($b_{\bm k n \tau}$) is the hole creation (annihilation) operator for the $n$th moir\'e valence band at momentum $\bm k$ and valley $\tau$. $\mathcal{E}_{\bm k}^{n\tau}$ is the single-particle band energy in the hole basis. 

\subsection{Mean-field calculation}
We theoretically calculate the interaction-driven QAHI state in $t$MoTe$_2$ at hole filling factor $\nu=1$, which has been experimentally realized\cite{Xiaodong2023a,Xiaodong2023b,Yihang2023_integer,Park2023b,observation_Xu2023}. In our calculation, we use a band-projected interacting Hamiltonian $\hat{\mathcal{H}}$ by retaining the top two moir\'e valence bands in $t$MoTe$_2$, which includes both the single-particle term $\hat{\mathcal{H}}_{\mathrm{KM}}$ of Eq.~\eqref{H0mor} and the Coulomb interaction term $\hat{\mathcal{H}}_\text{int}$ as follows,
\begin{equation}\label{H0Hint}
\begin{aligned}
\hat{\mathcal{H}}_\textrm{int}=\frac{1}{2} \sum V_{\bm k_1 \bm k_2 \bm k_3 \bm k_4}^{n_1 n_2 n_3 n_4}\left(\tau, \tau'\right)
b_{\bm k_1 n_1 \tau}^{\dagger} b_{\bm k_2 n_2 \tau'}^{\dagger} b_{\bm k_3 n_3 \tau'} b_{\bm k_4 n_4 \tau},
\end{aligned}
\end{equation}
where $V_{\bm k_1 \bm k_2 \bm k_3 \bm k_4}^{n_1 n_2 n_3 n_4}\left(\tau, \tau'\right)$ is the Coulomb potential $V_{\bm q}=2\pi e^2 \tanh{(|\bm q|d)}/(\epsilon |\bm q|)$ projected onto the moir\'e bands, where $d$ is the gate-to-sample distance and $\epsilon$ is the dielectric constant. In our calculation, we take $d=20$ nm as a representative value for the gate-to-sample distance, consistent with experimental setups—for example, as reported in Ref.~\cite{Xiaodong2023b}. For the dielectric constant, we use $\epsilon=15$, which is a typical phenomenological value adopted in studies of many-body effects in $t$MoTe$_2$ \cite{InteractionDriven2023,WangChong2024}. This phenomenological dielectric constant is intended to account both the screening from the device environment and internal screening contributions from remote electronic bands. 
The details on the construction of $\hat{\mathcal{H}}_\textrm{int}$ can also be found in Ref.~\cite{InteractionDriven2023}.

\begin{figure}[t]
\centering
\includegraphics[width=0.48\textwidth,trim=0 0 0 0,clip]{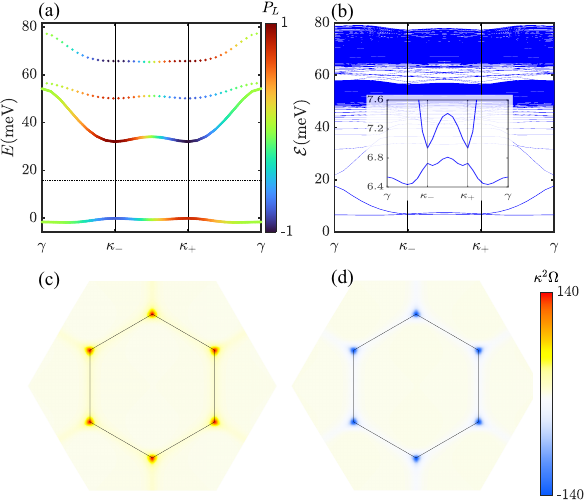}
\caption{(a) HF band structure of the QAHI at $\theta=3.5^\circ$ in the hole basis along high-symmetry path in the mBZ. Here $\gamma$ and $\kappa_{\pm}$ are, respectively, the center and corners of the mBZ. The solid (dashed) lines plot bands in $\tau=+ (-)$ valley. The dotted horizontal line marks the middle of the gap. The color represents the layer polarization $P_L$ with $+1$ $(-1)$ indicating the bottom (top) layer. (b) Magnon spectrum calculated for the $\nu=1$ QAHI state at $\theta=3.5^\circ$. The inset is a zoom-in plot. (c),(d) The Berry curvature $\Omega$ for the first and second band in (b). $\kappa$ is $4\pi/3a_M$. The black hexagon in (c) and (d) marks the mBZ.}
\label{fig2}
\end{figure}

We perform mean-field studies of $\hat{\mathcal{H}}$ using self-consistent HF approximation at $\nu=1$ to obtain the valley polarized QAHI state. By using Hartree-Fock decomposition, Eq.~\eqref{H0Hint} can be approximated as 
\begin{equation}
\begin{split}
\hat{\mathcal{H}}_{\mathrm{int}} \approx& \sum V_{\boldsymbol{k}_1 \boldsymbol{k}_2 \boldsymbol{k}_3 \boldsymbol{k}_4}^{n_1 n_2 n_3 n_4}\left(\tau, \tau^{\prime}\right)\left[\langle b_{\boldsymbol{k}_1 n_1 \tau}^{\dagger} b_{\boldsymbol{k}_4 n_4 \tau}\rangle b_{\boldsymbol{k}_2 n_2 \tau^{\prime}}^{\dagger}  b_{\boldsymbol{k}_3 n_3 \tau^{\prime}}\right.\\
&\left.
-\langle b_{\boldsymbol{k}_1 n_1 \tau}^{\dagger} b_{\boldsymbol{k}_3 n_3 \tau^{\prime}}\rangle b_{\boldsymbol{k}_2 n_2 \tau^{\prime}}^{\dagger} b_{\boldsymbol{k}_4 n_4 \tau}\right].
\end{split}
\end{equation}
For the QAHI state at $\nu=1$, both the momentum $\bm k$ and the valley index $\tau$ are good quantum numbers. Namely, the density matrix $\langle b_{\boldsymbol{k}_1 n_1 \tau}^{\dagger} b_{\boldsymbol{k}_3 n_3 \tau^{\prime}}\rangle$ is finite only when $\bm k_1=\bm k_3$ and $\tau=\tau'$. Therefore, the mean field Hamiltonian including both the single-particle kinetic energy and the Hartree-Fock approximation of $\hat{\mathcal{H}}_{\mathrm{int}}$ becomes
\begin{equation}
\begin{split}
\hat{\mathcal{H}}_{\mathrm{MF}} =& \sum \mathcal{E}_{\boldsymbol{k}}^{n \tau} b_{\boldsymbol{k} n \tau}^{\dagger} b_{\boldsymbol{k} n \tau}+\sum\left[V_{\boldsymbol{k}_1 \boldsymbol{k}_2 \boldsymbol{k}_2 \boldsymbol{k}_1}^{n_1 n_2 n_3 n_4}(\tau', \tau)\langle b_{\boldsymbol{k}_1 n_1 \tau'}^{\dagger} b_{\boldsymbol{k}_1 n_4 \tau'}\rangle\right.\\
&\left.
-V_{\boldsymbol{k}_1 \boldsymbol{k}_2 \boldsymbol{k}_1 \boldsymbol{k}_2}^{n_1 n_2 n_4 n_3}(\tau, \tau)\langle b_{\boldsymbol{k}_1 n_1 \tau}^{\dagger} b_{\boldsymbol{k}_1 n_4 \tau}\rangle\right] b_{\boldsymbol{k}_2 n_2 \tau}^{\dagger} b_{\boldsymbol{k}_2 n_3 \tau}.
\end{split}
\end{equation}
After self-consistent calculation, the mean field Hamiltonian $\hat{\mathcal{H}}_{\mathrm{MF}}$ can be formally written as 
\begin{equation}
\hat{\mathcal{H}}_{\text{MF}}=   \sum_{\boldsymbol{k}, \tau, \lambda} E_{\bm k}^{\lambda\tau} f_{\boldsymbol{k} \lambda \tau}^{\dagger} f_{\boldsymbol{k} \lambda \tau},
\end{equation}
where $E_{\bm k}^{\lambda\tau}$  and $\lambda$ are, respectively, the energy and band index for the interaction renormalized band structure. Here  $f_{\boldsymbol{k} \lambda \tau}^{\dagger}$ and $b_{\boldsymbol{k} n \tau}^{\dagger}$ operators are related by unitary transformations determined by the HF calculation.
In Fig.~\ref{fig2}(a), we show the mean-field band structure for the  QAHI state at $\theta=3.5^\circ$. The occupied band with a Chern number $C$ of $|C|=1$ is valley polarized and separated by a mean-field energy gap of 32 meV from unoccupied bands. 
The valley polarization in the QAHI state spontaneously breaks the time-reversal symmetry $\hat{\mathcal{T}}$, but does not break the continuous valley $U(1)$ symmetry.
For definiteness, the QAHI state is assumed to be polarized to the $\tau=+$ valley unless otherwise specified. 

\section{Topological magnons}\label{sec3}

\subsection{Magnon spectrum}
Based on the QAHI state in Fig.~\ref{fig2} (a),  we study intervalley collective excitations where the particle is excited from the occupied band at $\tau=+$ to the unoccupied bands at $\tau=-$. This excitation carries a single spin flip due to spin-valley locking and therefore, is equivalent to magnon.  The collective magnon excitations can be parametrized as \cite{Excitonband2015}
\begin{equation}\label{exct}
|\chi, \bm q\rangle=\sum_{\bm k, \lambda} z_{\bm k}^\lambda(\chi, \bm q) f_{\bm k+\bm q,\lambda, -}^{\dagger} f_{\bm k,1,+}|G\rangle,
\end{equation}
where $|G\rangle$ is the Slate-determinant ground state obtained in the HF approximation, $\bm q$ is the center-of-mass (CM) momentum of the magnon, and $\chi$ is an index that labels the magnon states at a given $\bm q$. The magnon state $|\chi, \bm q\rangle$ corresponds to a superposition of particle-hole pairs $f_{\bm k+\bm q,\lambda, -}^{\dagger} f_{\bm k,1,+}|G\rangle$ with the envelop function $z_{\bm k}^\lambda(\chi, \bm q)$, which satisfies the normalization condition $\sum_{\bm k, \lambda} |z_{\bm k}^\lambda(\chi, \bm q)|^2=1$. Variation of the energy $\langle \chi,\bm q| \hat{\mathcal{H}} |\chi,\bm q \rangle$ with respect to the parameter $z_{\bm k}^\lambda(\chi, \bm q)$ leads to the Bethe-Salpeter equation,
\begin{equation}\label{BS-equation1}
\begin{aligned}
&\mathcal{E}_{\chi,\bm q} z_{\bm k}^\lambda(\chi, \bm q)=\sum_{\bm p \lambda'}\mathcal{H}^{\lambda\lambda'}_{\bm k \bm p}(\bm q)z_{\bm p}^{\lambda'}(\chi, \bm q), \\
&\mathcal{H}^{\lambda\lambda'}_{\bm k \bm p}(\bm q)=(E_{\bm k+\bm q}^{\lambda -}- E_{\bm k}^{1 +})\delta_{\bm k \bm p}\delta_{\lambda\lambda'}
-\tilde{V}^{1\lambda\lambda'1}_{\bm p,\bm k+\bm q,\bm p+\bm q,\bm k}(+,-),
\end{aligned}
\end{equation}
where $\mathcal{H}^{\lambda\lambda'}_{\bm k \bm p}(\bm q)$ includes the quasiparticle energy cost of particle-hole transition as well as electron-hole attractive interaction. Here $\tilde{V}$ denotes Coulomb matrix element in the basis of $f_{\boldsymbol{k} \lambda \tau}^{\dagger}$ operators. The study of intravalley exctions (i.e., intravalley collective excitations) in $t$MoTe$_2$ can be found in Ref.~\cite{QuantumGeometry2024}.

We obtain the magnon energy $\mathcal{E}_{\chi,\bm q}$ and wave function $|\chi,\bm q \rangle$ at each $\bm q$ by numerically diagonalizing the matrix $\mathcal{H}^{\lambda\lambda'}_{\bm k \bm p}(\bm q)$. Numerical convergence tests of the magnon energies are presented in Appendix \ref{appa}. The calculated magnon
spectrum for the QAHIs phase at $\theta=3.5^\circ$ is shown in Fig.~\ref{fig2}(b) and has the following important features. (1) All the magnon excitations have finite energies with a minimum of 6.4 meV. This is consistent with the fact that the QAHI state does not break any continuous symmetry, and therefore, there is \textit{no} gapless Goldstone mode. The positive definiteness of the magnon excitation indicates the robustness of the QAHI state. (2) The lowest two magnon bands are isolated in energy from the continuous spectrum. (3)  There is a small gap of value 0.2 meV that separates the lowest two magnon bands at the mBZ corners $\bm{\kappa}_{\pm}$. [See the inset of Fig.~\ref{fig2}(b)].
We mainly focus on the first two magnon bands in the following. 

We further calculate the Berry curvatures $\Omega$ and Chern numbers $C$ of the first two magnon bands using the method of Ref.~\cite{KwanYves2021} (See also Appendix~\ref{appb} for details). The Berry curvatures $\Omega$, as shown in  Figs.~\ref{fig2}(c) and (d), are peaked at $\bm{\kappa}_{\pm}$ where the small gap is opened up and have opposite values for the first two magnon bands. The Chern numbers $C$, obtained from the integration of the Berry curvatures, are $-1$ and $+1$ for the first and second magnon bands, respectively. Therefore, the lowest two magnon bands are topological as characterized by the Chern numbers.

\begin{figure}[t]
\centering
\includegraphics[width=0.45\textwidth,trim=0 0 0 0,clip]{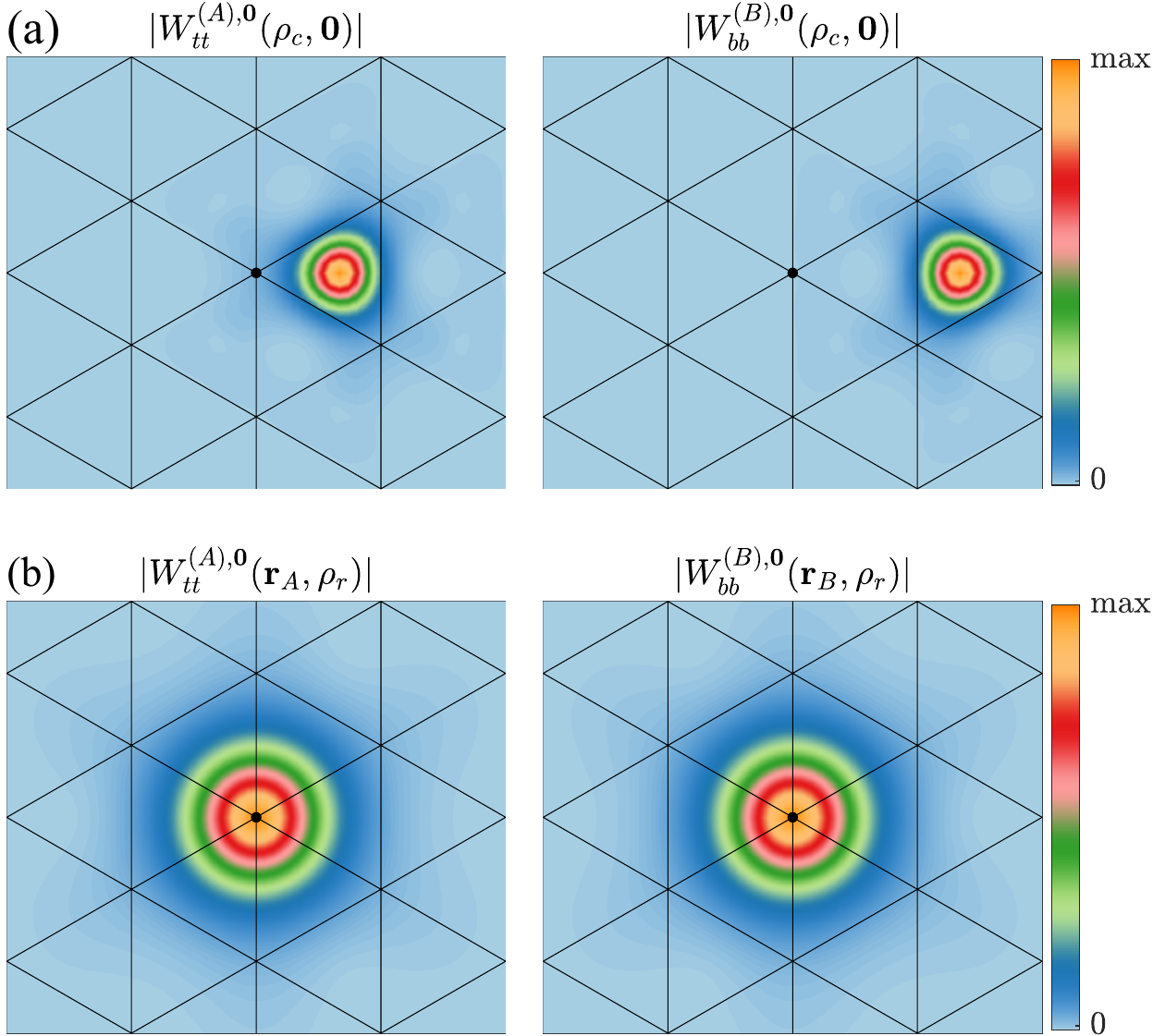}
\caption{The amplitude of magnon Wannier states for $\theta = 3.5^\circ$ at $\nu=1$. (a) $|W^{(A),\bm 0}_{tt}(\bm \rho_c,\bm \rho_r)|$ and $|W^{(B),\bm 0}_{bb}(\bm \rho_c,\bm \rho_r)|$ plotted in coordinate of $\bm \rho_c$ with $\bm \rho_r$ fixed at zero. (b) $|W^{(A),\bm 0}_{tt}(\bm \rho_c,\bm \rho_r)|$ and $|W^{(B),\bm 0}_{bb}(\bm \rho_c,\bm \rho_r)|$ plotted in coordinate of $\bm \rho_r$ with $\bm \rho_c$ fixed at $A$ and $B$ site, respectively. The black lines mark a triangular lattice with period $a_M$, and the black dot is the origin of the coordinates.}
\label{fig3}
\end{figure}

\subsection{Magnon tight-binding model}
To gain a deeper insight, we construct a tight-binding model based on magnon Wannier states for the first two magnon bands, which is feasible as their total Chern number is zero.  We start by defining the real-space magnon Bloch wavefunction for the $|\chi, \bm q\rangle$ state  as
\begin{equation}\label{rsmwf}
\begin{aligned}
g_{ll'}^{\chi,\bm q}(\bm r,\bm r')=\sum_{\bm k,\lambda}
z^{\lambda}_{\bm k}(\chi,\bm q)
\psi^{\lambda}_{\bm k+\bm q,l,-}(\bm r) [\psi^{1}_{\bm k,l',+}(\bm r')]^*,
\end{aligned}
\end{equation}
where $\bm r$ ($\bm r'$) is the in-plane position of the $p^*$ ($h^*$) particle, and $l$ ($l'$) is the corresponding layer index. Here $p^*$ ($h^*$)  labels the particle (hole) in the hole basis that we employ. $\psi^{\lambda}_{\bm k+\bm q,l,-}(\bm r)$ [$\psi^{1}_{\bm k,l',+}(\bm r')$] is the quasiparticle Bloch wavefunction at valley $\tau=-(+)$ obtained by the mean-field calculation. In contrast to the electron Bloch wavefunction, the magnon Bloch state describes a two-particle state with two coordinates $\bm r$ and $\bm r'$. For convenience, we also define the CM coordinate  $\bm\rho_c=(\bm r+\bm r')/2$ and the relative coordinate $\bm\rho_r=\bm r-\bm r'$.

To capture the first two magnon bands, we need two magnon Wannier states, which can be formally constructed as~\cite{MaximallyHaber2023,ExcitonicJankowski2024,InteractionDavenport2024},
\begin{equation}\label{wsamlv}
W_{ll'}^{(\alpha),\bm R}(\bm \rho_c,\bm \rho_r)=\frac{1}{\sqrt{N}}\sum_{\bm q} e^{i\bm q \bm R}   W_{ll'}^{(\alpha),\bm q}(\bm \rho_c,\bm \rho_r),
\end{equation}
\begin{equation}\label{wlcbnt}
\begin{aligned}
W_{ll'}^{(\alpha),\bm q}(\bm \rho_c,\bm \rho_r)=
\sum_{\chi=1,2}C^{(\alpha)}_{\chi,\bm q}
g_{ll'}^{\chi,\bm q}(\bm \rho_c+\frac{\bm \rho_r}{2},\bm \rho_c-\frac{\bm \rho_r}{2}).
\end{aligned}
\end{equation}
In Eq.~\eqref{wsamlv}, $W^{(\alpha),\bm R}(\bm \rho_c,\bm \rho_r)$ is the magnon Wannier state with the CM wavefunction localized in the moir\'e unit cell $\bm R$, $W^{(\alpha),\bm q}(\bm \rho_c,\bm \rho_r)$ is the corresponding Bloch-like state derived from the Wannier state, and $N$ is the number of $\bm q$ points included in the summation. The index $\alpha$ labels the two magnon Wannier states, and we use the $A$ and $B$ symbols for the two values of $\alpha$ with reasons to be clear shortly. In Eq.~\eqref{wlcbnt}, the Bloch-like state $W^{(\alpha),\bm q}(\bm \rho_c,\bm \rho_r)$ is obtained by a unitary transformation from the magnon Bloch wavefunctions, where the coefficients $C^{(\alpha)}_{\chi,\bm q}$ form a unitary matrix $U(\bm q)$ at each $\bm q$,
\begin{equation}
U(\bm q)=\left[
\begin{array}{cc}
C^{(A)}_{1,\bm q} & C^{(B)}_{1,\bm q} \\
C^{(A)}_{2,\bm q} & C^{(B)}_{2,\bm q} \\
\end{array}
\right].
\label{Umagnon}
\end{equation}

We now provide a procedure to determine the unitary matrix $U(\bm q)$. Since the fermionic model in Eq.~\eqref{thhtn} is based on a honeycomb lattice formed by $A$ and $B$ sites, we can take an ansatz that the CM wavefunction of the magnon Wannier state  $W^{(\alpha),\bm R}(\bm \rho_c,\bm \rho_r)$ is centered at the $\alpha$ site for $\alpha= A, B$. For this purpose, we define the following sublattice polarization function, 
\begin{equation}\label{Fpolarizaiton}
\begin{aligned}
F^{(\alpha)}_{\bm q}=|W_{tt}^{(\alpha),\bm q}(\bm r_A,\bm 0)|^2-|W_{bb}^{(\alpha),\bm q}(\bm r_B,\bm 0)|^2,
\end{aligned}
\end{equation}
where $\bm r_A=(a_M/\sqrt{3},0)$ and $\bm r_B=(2a_M/{\sqrt{3}},0)$ are, respectively, the positions of $A$ and $B$ sites at the $\bm R=\bm 0$ unit cell. Under the normalization constraint of $\sum_{\chi=1,2}|C^{(\alpha)}_{\chi,\bm q}|^2=1$, we obtain $C^{(\alpha)}_{\chi,\bm q}$ by maximizing (minimizing) the function $F^{(\alpha)}_{\bm q}$ for $\alpha=A (B)$. This procedure automatically leads to a unitary matrix $U(\bm q)$. We further fix the gauge by requiring that $W_{tt}^{(A),\bm q}(\bm r_A,\bm 0)>0$ and $W_{bb}^{(B),\bm q}(\bm r_B,\bm 0)>0$. Here we employ the locking between site and layer, as the $A$ ($B$) site is polarized to the $t$ ($b$) layer.

 The constructed $W^{(\alpha),\bm R}(\bm \rho_c,\bm \rho_r)$ 
using the above method are shown in Fig.~\ref{fig3} for $\bm R=\bm 0$ and $\theta=3.5^\circ$, which plots the amplitude with fixed $\bm \rho_r$ in Fig.~\ref{fig3}(a) and fixed $\bm \rho_c$ in Fig.~\ref{fig3}(b), respectively. Here, we only show $W_{tt}^{(A),\bm 0}(\bm \rho_c,\bm \rho_r)$ and $W_{bb}^{(B),\bm 0}(\bm \rho_c,\bm \rho_r)$ since the two magnon Wannier states are mostly distributed at $l=l'=t$ and $l=l'=b$, respectively. Consistent with the initial ansatz, the CM wavefunction of $W^{(\alpha),\bm R}(\bm \rho_c,\bm \rho_r)$ are indeed centered at the $\alpha$ site. Meanwhile, the relative motion wavefunction of $W^{(\alpha),\bm R}(\bm \rho_c,\bm \rho_r)$, as shown in Fig.~\ref{fig3}(b), is $s$-wave like, which demonstrates the particle-hole bound state in the magnon. Overall, the magnon Wannier states capture the local spin flip at sites $A$ and $B$.

A tight-binding Hamiltonian can be constructed for the magnons based on the Wannier states,
\begin{equation}\label{hopping_main}
\begin{split}
H_{\text{mag}}&=\sum_{\alpha\beta}\sum_{\bm R\bm R'}T_{\alpha\beta}(\bm R'-\bm R) a^\dagger_{\bm R\alpha}a_{\bm R'\beta},\\
T_{\alpha\beta}(\bm R)&=\frac{1}{N}\sum_{\bm q \chi} e^{i\bm q \bm R} U^{\dagger}_{\alpha\chi}(\bm q) \mathcal{E}_{\chi,\bm q} U_{\chi\beta}(\bm q),
\end{split}
\end{equation}
where $a^\dagger_{\bm R\alpha}$ ($a_{\bm R'\beta}$) is the magnon creation (annihilation) operator at site $\alpha$ ($\beta$),
$\mathcal{E}_{\chi,\bm q}$ is the magnon energy obtained from Eq.~\eqref{BS-equation1}, and $U(\bm q)$ is the unitary matrix defined in Eq.~\eqref{Umagnon}.

\begin{figure}[t]
\centering
\includegraphics[width=0.48\textwidth,trim=0 0 0 0,clip]{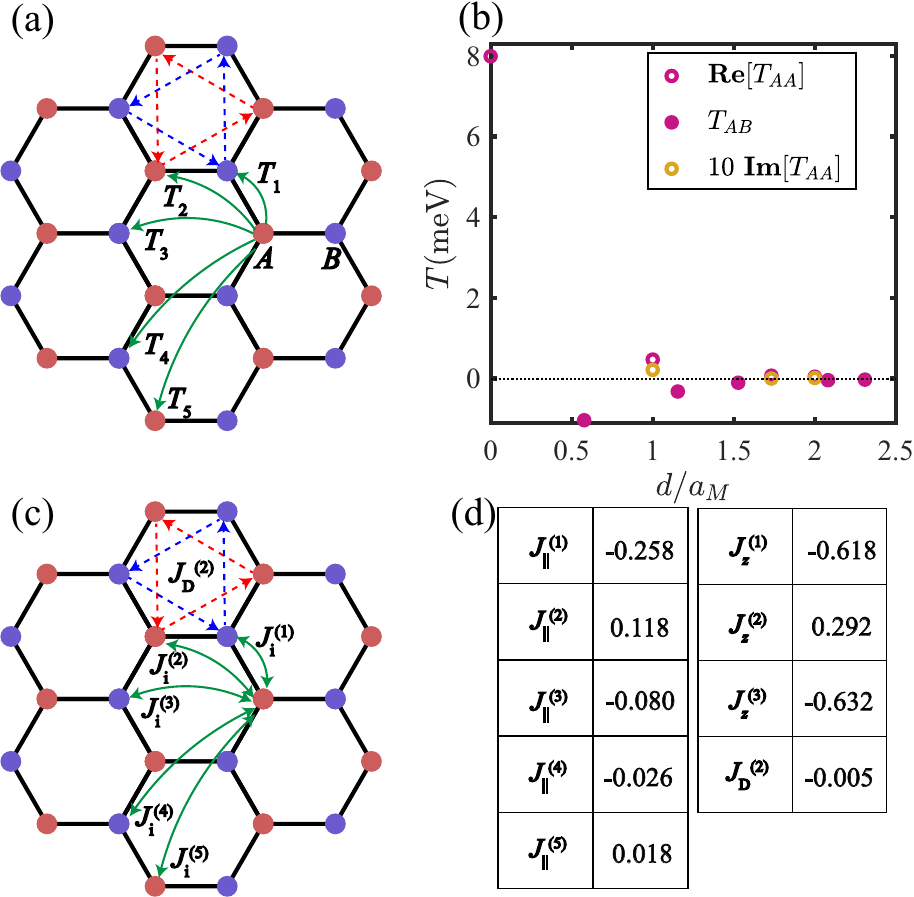}
\caption{(a) Illustration of hopping parameters in the magnon tight-binding model. The next-nearest neighbor hoppings have a sublattice and direction-dependent phase factors, as illustrated by the dashed arrows. (b) Numerical values of hopping terms in the magnon tight-binding model as a function of two-site distance $d$. (c) Illustration of the effective spin model on the honeycomb lattice. The dashed arrows indicate the DMI between the next-nearest neighbors. (d) Numerical values of spin coupling constants in the unit of meV at $\theta=3.5^\circ$.}
\label{fig4}
\end{figure}

In Fig.~\ref{fig4}(b), we plot $T_{\alpha\beta}(\bm R)$ as a function of the hopping distance $d=|\bm R+\bm r_\beta-\bm r_\alpha|$, where  the hopping $T_{\alpha\beta}(\bm R)$  decreases exponentially with $d$ for large $d$ \cite{MaximallyHaber2023}.  Notably, the next-nearest neighbor hopping terms are complex with sublattice and direction-dependent phase factors  $e^{i\phi_T \nu_{ij}}$, where $\phi_T$ is a phase and $\nu_{ij}=+1 (-1)$ if the hopping from the site $i$ to $j$ is along (against) the dashed arrows in Fig.~\ref{fig4}(a). Therefore, Eq.~\eqref{hopping_main} realizes an effective Haldane model for magnons \cite{Haldane1988,KimRealization2016,Owerre_first2016, McClartyTopological2022}, where the complex next-nearest neighbor hopping terms induces topological gap at $\bm \kappa_{\pm}$, endowing the two magnon bands with opposite Chern numbers. We emphasize that the onsite energy $T_{\alpha \alpha} (\bm 0) \approx 8$ meV is an important term, as the magnon energy is already defined relative to the ground state.

\begin{figure}[t]
\centering
\includegraphics[width=0.47\textwidth,trim=0 0 0 0,clip]{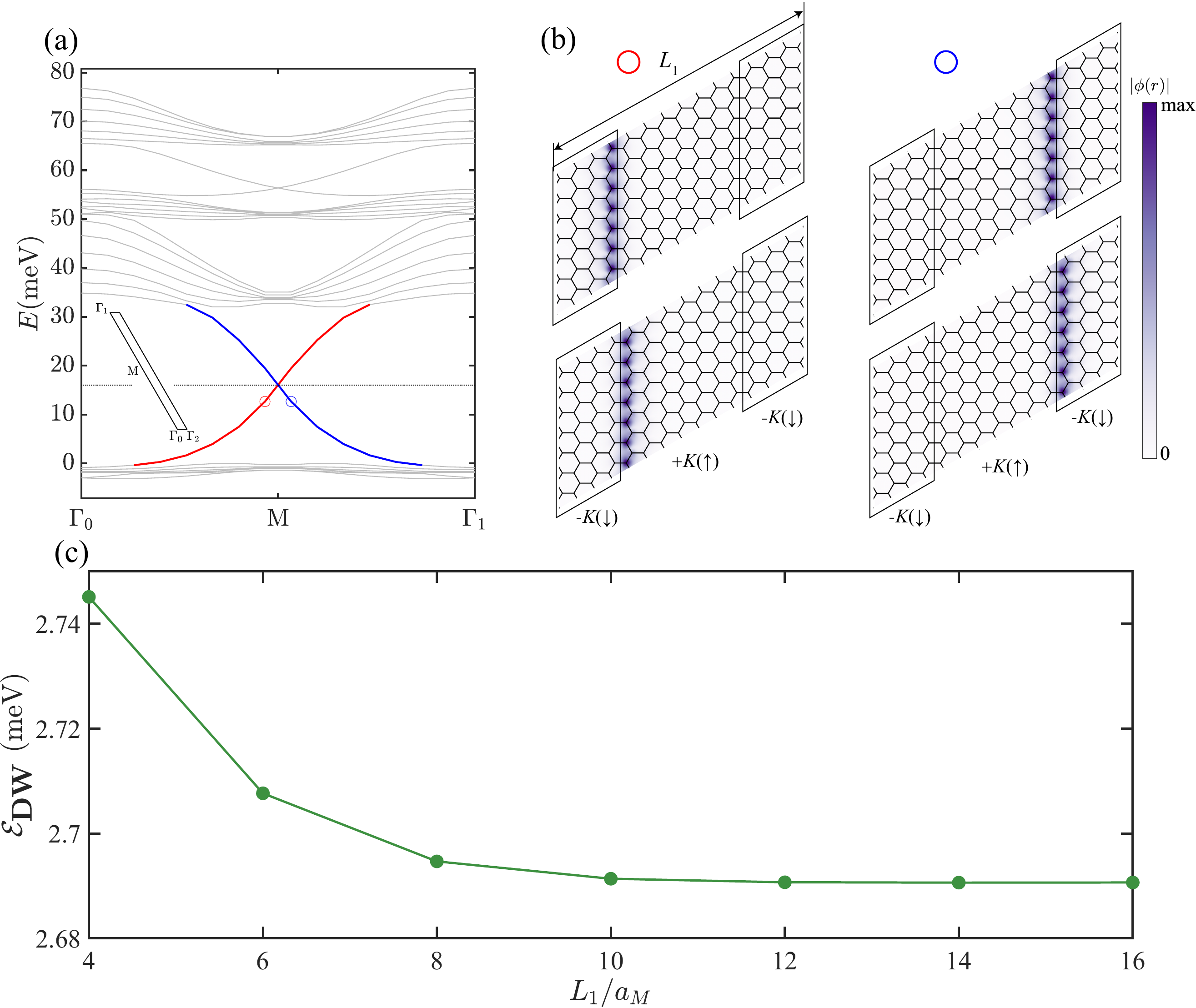}
\caption{(a) HF band structure of the domain structure
shown in (b) with $L_1=14 a_M$, where the valley polarizations change signs twice along the $\hat{\bm e}_1=(\sqrt{3}/2,1/2)$ direction. The inset shows the Brillouin zone (bz) of the domain superstructure. The gray bands belong to the bulk states. The red (blue) line indicates doubly degenerate in-gap chiral states that co-propagate along the left (right) domain wall. The horizontal dotted line marks the middle of the insulating gap of the uniform QAHI at $\nu=1$. (b) The wavefunction amplitude of chiral edge states labeled on the spectrum in (a). (c) The domain wall energy as a function of $L_1$. These results are calculated at $\theta=3.5^\circ$.}
\label{fig5}
\end{figure}

\section{Domain walls}\label{sec4}
We now study another type of excited state, i.e., domain walls that separate real-space regions with opposite valley (spin) polarizations. We study a domain structure shown in Fig.~\ref{fig5}(b), where the valley polarizations change signs twice along the $\hat{\bm e}_1=(\sqrt{3}/2,1/2)$ direction, leading to two parallel domain walls along the  $\hat{\bm e}_2=(0,1)$ direction. We apply periodic boundaries for the superstructure with a length $L_i=N_i a_M$ along the $\hat{\bm e}_i$ direction for $i=1,2$. We note that the superstructure has a period of $L_1$ along the $\hat{\bm e}_1$ direction due to the presence of the domains, but $a_M$ along the $\hat{\bm e}_2$ direction. Therefore, the  Brillouin zone for the superstructure (see inset of Fig.~\ref{fig5}(a) for illustration), denoted as bz for shorthand notation, is reduced compared to the mBZ.

We perform self-consistent HF calculation for the superstructure with two domain walls by sampling the bz using a centered scheme with a $1\times N_2$ $k$-point mesh, where $N_2$ is taken to be 15. In the iterative HF calculation, we start with a configuration with two domain walls, which remain after the self-consistent calculation. In the iteration, the filling factor is fixed at $\nu=1$. 
The resulting mean-field band structure is shown in Fig.~\ref{fig5}(a) for a superstructure with $L_1= 14 a_M$. Since each domain wall separates Chern insulators with opposite valley polarization and therefore, opposite Chern numbers, it binds chiral electronic edge states within the bulk insulating gap. Because the change of Chern numbers across each domain wall is 2, there are two chiral electronic edge states per domain wall. Moreover, the left and right domain walls host chiral edge states with opposite velocities. All these expected features are indeed obtained from our calculation, as shown by the energy spectrum in   Fig.~\ref{fig5}(a) and the chiral state wavefunctions in Fig. \ref{fig5}(b). We view these domain walls as topological, as they host the chiral states.

The domain walls are excited states and cost a finite energy compared to the ground state. We denote the energy cost of a domain wall per unit cell along the $\hat{\bm e}_2$ direction as $\mathcal{E}_{\text{DW}}$. Here $\mathcal{E}_{\text{DW}}=(E_2-E_0)/(2N_2)$, where $E_2$ and $E_0$ are respectively the total energy of the $L_1\times L_2$ superstructure with two and no domain walls. In Fig.~\ref{fig5}(c), we plot $\mathcal{E}_{\text{DW}}$ as a function of $L_1$. The convergence of $\mathcal{E}_{\text{DW}} \approx 2.69 $ meV is achieved for  $L_1 = 14 a_M$, above which the interactions between the two domain walls separated by a length of $L_1/2$ is negligible.

\section{Effective spin model}\label{sec5}
We propose an effective spin model to provide a unified low-energy description of the magnon and domain wall excitations. The construction of this effective model is informed by the following properties of the system. (1) The ground state is a ferromagnetic state with aligned out-of-plane spin polarizations on a honeycomb lattice.  (2) The two low-energy magnon bands are derived from spin flips on the $A$ and $B$ sublattices of the honeycomb lattice. (3) The system hosts Ising-type domain walls.  We use the following nonlinear sigma model defined on a honeycomb lattice for the effective spin model, with the Lagrangian given by,
\begin{equation}\label{knbeph_main1}
\begin{split}
\mathcal{L}_S&=\mathcal{B}_S-\mathcal{E}[\bm m],\\
\mathcal{B}_S&=
-\frac{\hbar}{2}\sum_{\bm R\alpha}n_\alpha \mathcal{\bm A}[\bm m_{\bm R\alpha}]\cdot \partial_t\bm m_{\bm R\alpha}, \\
\mathcal{E}[\bm m]&
=\sum'_{\bm R \bm R'\alpha\beta}\sum_{i} J^{\alpha\beta}_{i}(\bm R'-\bm R)
m^i_{\bm R\alpha}m^i_{\bm R'\beta}\\
&+\sum'_{\bm R \bm R'\alpha}J^{\alpha\alpha}_{D}(\bm R'-\bm R)(\bm m_{\bm R\alpha} \times \bm m_{\bm R'\alpha})\cdot \hat{z} ,
\end{split}
\end{equation}
where $\bm m_{\bm R\alpha}=(m^x_{\bm R\alpha},m^y_{\bm R\alpha}, m^z_{\bm R\alpha})$ is a unit vector that describes the magnetization direction at sublattice $\alpha$ and unit cell $\bm R$ of a honeycomb lattice. Here $\mathcal{B}_S$ is the kinetic Berry phase, where $\mathcal{\bm A}[\bm m_{\bm R\alpha}]$ is the effective spin gauge field defined by $\bm{\nabla}_{\bm m} \times \mathcal{\bm A}[\bm m]=\bm m$, $n_\alpha$ is the average occupation number at the sublattice $\alpha$, and $\partial_t$ is the derivative with respect to time $t$. At $\nu=1$, $n_\alpha$ is $1/2$. The energy functional $\mathcal{E}[\bm m]$ is expressed by the classical spin couplings on the honeycomb lattice, of which the form is constrained by the symmetry of the system, in particular the $\hat{D}_{3d}$ point group symmetry and the spin $U(1)$ symmetry. The first line in the expression of $\mathcal{E}[\bm m]$ describes the spin couplings in the XXZ Heisenberg model, where the in-plane coupling constant satisfies $J^{\alpha\beta}_{x}(\bm R)=J^{\alpha\beta}_{y}(\bm R)=J^{\alpha\beta}_{\|}(\bm R)$ due to the spin $U(1)$ symmetry and $J^{\alpha\beta}_{z}(\bm R)$ is the out-of-plane coupling constant. The second line in $\mathcal{E}[\bm m]$ is the DMI between spins on next-nearest neighbors, with the direction from site $\bm R \alpha$ to $\bm R' \alpha$ specified by the dashed arrows in Fig.~\ref{fig4}(c). The prime in the summations denotes that each bond is counted only once. The form of $\mathcal{L}_S$ in Eq.~\eqref{knbeph_main1} is justified by a microscopic derivation presented in Appendix \ref{appc}.

The coupling constants $\{J_{\|}, J_z, J_D\}$ of the above effective model can be extracted from the numerical results of magnons in Sec.~\ref{sec3}  and domain walls in Sec.~\ref{sec4}.
We first consider magnon excitations by taking the in-plane components $\{m^x_{\bm R\alpha}, m^y_{\bm R\alpha} \}$ as small fluctuations. With the approximation of $m^z_{\bm R\alpha}\approx 1- \frac{1}{2}[(m^x_{\bm R\alpha})^2+(m^y_{\bm R\alpha})^2]$, the energy functional $\mathcal{E}[\bm m]$ is expanded to second order in $\{m^x_{\bm R\alpha}, m^y_{\bm R\alpha} \}$, while the kinetic Berry phase can be expressed as, 
\begin{equation}\label{kinetic_Berry}
\begin{split}
\mathcal{B}_S&=
-\frac{\hbar}{4}\sum_{\alpha}\sum_{\bm R}n_\alpha( m^{x}_{\bm R{\alpha}}\partial_t m^{y}_{\bm R{\alpha}}-
m^{y}_{\bm R{\alpha}}\partial_t m^{x}_{\bm R{\alpha}}),
\end{split}
\end{equation}
We derive the equation of motion for $m^{+}_{\bm R{\alpha}}=m^{x}_{\bm R{\alpha}}+i m^{y}_{\bm R{\alpha}}$ using the Lagrangian $\mathcal{L}_S$ and compare it with that for the magnon annihilation operator  $a_{\bm R{\alpha}}$ using the magnon tight-binding Hamiltonian in Eq.~\eqref{hopping_main}.  
The comparison is motivated by the Holstein–Primakoff transformation that maps the spin operators to boson operators and relates the spin coupling constants with the hopping parameters in Eq.~\eqref{hopping_main},
\begin{align}
&\sum_{\bm R\beta}J^{\alpha\beta}_z(\bm R)=-\frac{1}{4} T_{\alpha \alpha}(\bm 0), \label{JzT0}\\
&J^{\alpha\beta}_{\|}(\bm R)=\frac{1}{4}\text{Re}[T_{\alpha\beta}(\bm R)], \label{J||}\\
&J^{\alpha\alpha}_{D}(\bm R)=-\frac{1}{4}\text{Im}[T_{\alpha\alpha}(\bm R)]. \label{JD}
\end{align}
This set of equations shows that coupling constants $J_{\|}$ and $J_D$ are directly determined by the hopping parameters, while only a sum rule is obtained for the out-of-plane coupling constants $J_z$. Equation~\eqref{JD} demonstrates that the DMI coupling constant $J^{\alpha\alpha}_{D}(\bm R)$ is proportional to $\text{Im} [T_{\alpha\alpha}(\bm R)]$. Since $\text{Im} [T_{\alpha\alpha}(\bm R)]$ is finite only for next-nearest neighbors as shown in Fig.~\ref{fig4}(a), we only keep the DMI terms between next-nearest neighbors in the effective spin model. We list numerical values of  $J_{\|}$ and $J_D$  in Fig.~\ref{fig4}(d). The in-plane coupling constants between the nearest neighbors, second and third are, respectively, $J_{\|}^{(1)}=-0.258$ meV, $J_{\|}^{(2)}=0.118$ meV, and $J_{\|}^{(3)}=-0.08$ meV. For comparison, the DMI coupling constant between the second-nearest neighbors is $J_D^{(2)}=-0.005$ meV, which is orders of magnitude smaller than the corresponding $J_{\|}^{(2)}$.

Although small, $J_D^{(2)}$ can not be treated as negligible, as it directly determines the topological gap $\Delta_g$ between the first and second magnon bands at the mBZ corners as follows,
\begin{equation}
\Delta_g=6 \sqrt{3}\left|\operatorname{Im} [T_2]\right|=24 \sqrt{3}\left|J_D^{(2)}\right|\approx0.2\ \textrm{meV},
\end{equation}
where $T_2$ is the magnon hopping parameter between the second-nearest neighbors on the honeycomb lattice. Because of the prefactor  $24 \sqrt{3}$, $\Delta_g$ and $J_D^{(2)}$ differ by two orders of magnitude.

To further determine the value of each $J^{\alpha\beta}_{z}(\bm R)$, we turn to the domain wall excitation. We consider a domain wall along the $\hat{\bm e}_2$ direction that separates two regions with opposite out-of-plane magnetization, i.e., $m_{\bm R \alpha}^{z}=-1$ and $m_{\bm R' \alpha'}^{z}=+1$, where  $\bm R \alpha$ ($\bm R' \alpha'$) are sites on the left (right) hand side of the domain wall. The energy cost per unit cell along $\hat{\bm e}_2$ of the domain wall compared to the ferromagnetic ground state is
\begin{equation}\label{energy_domain}
\begin{split}
\mathcal{E}_{\text{DW}}=-2J^{(1)}_z-8J^{(2)}_z-6J^{(3)}_z.
\end{split}
\end{equation}
Here $J^{\alpha\beta}_{z}(\bm R)$ is written as the $n$th nearest-neighbor terms $J^{(n)}_{z}$, which are truncated up to $n=3$. From the numerical calculation presented in Sec.~\ref{sec4}, we have  $\mathcal{E}_{\text{DW}} \approx 2.69$ meV. 
In addition to Eqs.~\eqref{JzT0} and ~\eqref{energy_domain}, another equation can be derived by comparing the energy difference per unit cell between an out-of-plane antiferromagnetic (AF$_z$) state and the ferromagnetic QAHI ground state as,
\begin{equation}\label{linear_eqs}
\begin{split}
E_{\text{AF}_z}-E_{\text{QAHI}}=-6J^{(1)}_z-6J^{(3)}_z.
\end{split}
\end{equation}
Here the AF$_z$ state has opposite out-of-plane spin polarizations on the $A$ and $B$ sublattices and is a meta-stable state that can be obtained within the self-consistent HF calculation, which leads to $E_{\text{AF}_z}-E_{\text{QAHIs}}=7.5$ meV. By combining Eqs.~\eqref{JzT0}, ~\eqref{energy_domain} and ~\eqref{linear_eqs}, we obtain the three out-of-plane coupling constants $[J^{(1)}_z, J^{(2)}_z, J^{(3)}_z]=(-0.62, 0.29, -0.63)$ meV. Here $J^{(1)}_z $ is negative, indicating the ferromagnetic coupling nature between the nearest neighbors on the honeycomb lattice. The magnitude of $J^{(n)}_z$ is larger than that of  $J^{(n)}_{\|}$, which is a manifestation of the Ising anisotropy.

The effective spin model in Eq.~\eqref{knbeph_main1} is now fully determined with the coupling constants obtained from microscopic calculations. This effective model gives rise to the correct out-of-plane ferromagnetic ground state, and quantitatively reproduces the topological magnon spectrum and the energy cost of domain wall excitations.

In our analysis, we neglect the magnetic dipole-dipole interaction, which could play a role in the formation of magnetic domains. However, its energy scale is significantly smaller than the dominant energies in moir\'e systens.  For $t$MoTe$_2$ with  $\theta=3.5^\circ$, the moir\'e period $a_M$ is about 5.8 nm. As an order-of-magnitude estimate, we consider two electrons with out-of-plane spin polarization separated by an in-plane distance $a_M$, of which the magnetic dipole-dipole interaction strength is $\frac{\mu_0 \mu_B^2}{4\pi a_M^3}\approx 2.8\times 10^{-7}$ meV, which is negligible than the energy scale that we focus on. Here $\mu_0$ is the vacuum magnetic permeability and $\mu_B$ is the Bohr magneton.

We can estimate the magnetic ordering temperature for the ferromagnetic ground state based on the effective spin model. Here, we use a mean-field theory to reveal the physics. The mean-field energy for a single site is  approximated as, 
\begin{equation}
\mathcal{E}_{\mathrm{MF}}=-h m^z, \quad h=-\left\langle m^z\right\rangle \sum_{\boldsymbol{R} \beta} J_z^{\alpha \beta}(\boldsymbol{R}),
\end{equation}
where $h$ is the effective magnetic field generated by neighboring spins. The thermal average value $\left\langle m^z\right\rangle$ is then  given by,
\begin{equation}
\begin{split}
\left\langle m^z\right\rangle=&\frac{\int d \boldsymbol{m}\ m^z e^{-\frac{\mathcal{E}_{\mathrm{MF}}}{k_B T}}}{\int d \boldsymbol{m} e^{-\frac{\mathcal{E}_{\mathrm{MF}}}{k_B T}}}=\coth \left(\frac{h}{k_B T}\right)-\frac{k_B T}{h} 
\approx \frac{1}{3} \frac{h}{k_B T},
\end{split}
\end{equation}
where $k_B$ is the Boltzmann constant, the integration over $\bm m$ is performed on the unit sphere to account for its three-dimensional nature, and the final expression holds in the small-$h$ limit. Therefore, a mean-field critical temperature is 
\begin{equation}
T_c \approx-\frac{\sum_{\bm R \beta} J_z^{\alpha \beta}(\boldsymbol{R})}{3 k_B}=\frac{T_{\alpha \alpha}(0)}{12 k_B} \approx 7.7 \mathrm{~K}.
\end{equation}
Experimentally, the Curie temperature for the QAHI state in $t$MoTe$_2$ with $\theta$ around $3.5^\circ$ is about 12 K ~\cite{Xiaodong2023a,Xiaodong2023b,observation_Xu2023}. Our mean-field estimation of $T_c$ is qualitatively consistent with this experimental value. 

We can employ an alternative method to estimate the magnetic ordering temperature based on magnon occupation. Here, each magnon corresponds to a single spin flip and has an occupation number that follows the Bose-Einstein distribution $f\left(\mathcal{E}_{\chi, \boldsymbol{q}}\right)=(e^{\frac{\mathcal{E}_{\chi, \boldsymbol{q}}}{k_B T}}-1)^{-1}$, where $\mathcal{E}_{\chi, \boldsymbol{q}}$ is the magnon energy. Using the relation $\left\langle m^z\right\rangle=\frac{1}{2}-\frac{1}{N} \sum_{\chi, \boldsymbol{q}} f\left(\mathcal{E}_{\chi, \boldsymbol{q}}\right)$, we estimate the transition temperature to be $T_c\approx53.5$ K by requiring $\left\langle m^z\right\rangle=0$. Here, $N$ is the number of $\bm q$ points in the summation.

Although both estimates are approximate, their comparison suggests that domain wall thermal proliferation—rather than magnon thermal population—are the primary factor limiting the transition temperature. Here, we would like to note that our intention is to offer an order-of-magnitude estimate rather than a precise prediction. A more accurate determination of the transition temperature would require methods such as classical or quantum Monte Carlo analysis, which is beyond the scope of this work.

\section{discussion and summary}\label{sec6}
In summary, we present a theoretical study of magnon and domain wall excitations for the QAHI state in twisted bilayer $t$MoTe$_2$ at $\nu=1$. Our numerical calculation starts from a generalized interacting Kane-Mele model projected onto the first two moir\'e bands. Based on the mean-field QAHI ground state and the Bethe-Salpeter equation, we obtain the magnon spectrum, where the first two bands have opposite Chern numbers. An effective Haldane model is constructed to characterize the two magnon bands. We further study a structure of domain walls that separate
regions with opposite valley polarization and Chern numbers. The energy cost of the domain wall is calculated. The above two magnetic excitations are phenomenologically described by an effective spin model, which consists of in-plane and out-of-plane Heisenberg spin coupling, as well as the DMI terms between next-nearest neighbors on the honeycomb lattice. The DMI terms are crucial to account for the magnon topology. The values of coupling constants in the effective spin model are determined from the numerical results for magnons and domain walls. The estimation of magnetic ordering temperature $T_c$ based on the spin coupling constants is qualitatively consistent with the experimental value. We note that the $k_B T_c \approx 0.7$ meV estimated from the effective spin model is an order of magnitude smaller than the charge gap $(\sim 32~\text{meV})$ obtained from the HF band structure [Fig.~\ref{fig2}(a)], and also a few times smaller than the minimum magnon excitation energy $(\sim 6.4~\text{meV})$ [Fig.~\ref{fig2}(b)]. This indicates that the domain-wall thermal proliferation limits the Curie temperature. 

The QAHI ground state does \textit{not} allow the description in terms of the occupation of a set of exponentially localized orbitals on a periodic lattice due to Wannier obstruction. Nevertheless, we show that low-energy magnon bands on top of the QAHI ground state can have a tight-binding model description based on localized magnon Wannier states. It should be noted that the magnons are two-particle bound states of electron and hole, which involves both occupied and unoccupied states. Given the Wannier obstruction of the QAHI ground state, we propose using the Lagrangian in Eq.~\eqref{knbeph_main1} to describe the effective spin model, which incorporates the fact that the average fermion occupation number per site is fractional rather than an integer. This contrasts with the effective spin Hamiltonian of a Mott insulator, where electrons are localized at each site with integer occupation. Due to the magnon topology, we construct a lattice-based nonlinear sigma model instead of the conventional continuous nonlinear sigma model, which has been used to describe the long-wavelength behavior of collective excitations\cite{SpontaneousMoon1995,WuFengcheng2020Collective,WuFengcheng2020Quantum,KhalafSoft2020,TaigeDiverse2023}. The continuous model discards lattice information, making it unsuitable for capturing the topology of collective excitations. 

$t$MoTe$_2$ serves as a representative system because it realizes an interacting Kane-Mele model on an effective honeycomb lattice. In this system, the underlying lattice model is particularly well-defined, enabling valuable insights into topological magnon excitations. In particular, it allows for a detailed analysis based on magnon Wannier states and effective hopping models. Our results demonstrate that the topology of electronic bands can be inherited by collective excitations (see Appendix~\ref{appc} for more discussion), suggesting that topological magnons could be a general feature in QAHIs and other topological phases. Furthermore, the methodology developed here—analyzing topological excitations through Wannier-based magnon models—can be extended to a broad class of systems. Several moir\'e materials—including those based on graphene and TMDs—are known to host QAHI phases with spontaneous ferromagnetism ~\cite{Cao2018,Cao2018a,Sharpe2019,Serlin2020,Chen2020,Polshyn2020,TschirhartImaging2021,Stepanov2021,Grovermosaic2022,LiQuantum2021}. Topological collective excitations have already been explored in systems such as twisted bilayer graphene \cite{KwanYves2021}.
In this regard, it would be interesting to revisit the collective excitations of interaction-driven symmetry-breaking insulating states in magic-angle twisted bilayer graphene and explore whether a lattice formulation of the nonlinear sigma model can be developed \cite{KumarAjesh2021}.

\section{ACKNOWLEDGMENTS}
We thank Xun-Jiang Luo for the valuable discussions. This work is supported by National Key Research and Development Program of China (Grants No. 2022YFA1402400 and No. 2021YFA1401300), National Natural Science Foundation of China (Grant No. 12274333 and No. 12404084). W.-X. Q. is also supported by the China Postdoctoral Science Foundation (Grants No. 2024T170675 and No. 2023M742716). The numerical calculations in this paper have been done on the supercomputing system in the Supercomputing Center of Wuhan University.

\textit{Note Added}. As this manuscript was being prepared, we learned of a related study by W.-T. Zhou et al. \cite{ItinerantZhou2025}, which reaches consistent conclusions about the magnon topology in $t$MoTe$_2$.
\appendix

\section{Numerical Convergence}\label{appa}

\begin{figure}[t]
\centering
\includegraphics[width=0.50\textwidth,trim=0 0 0 0,clip]{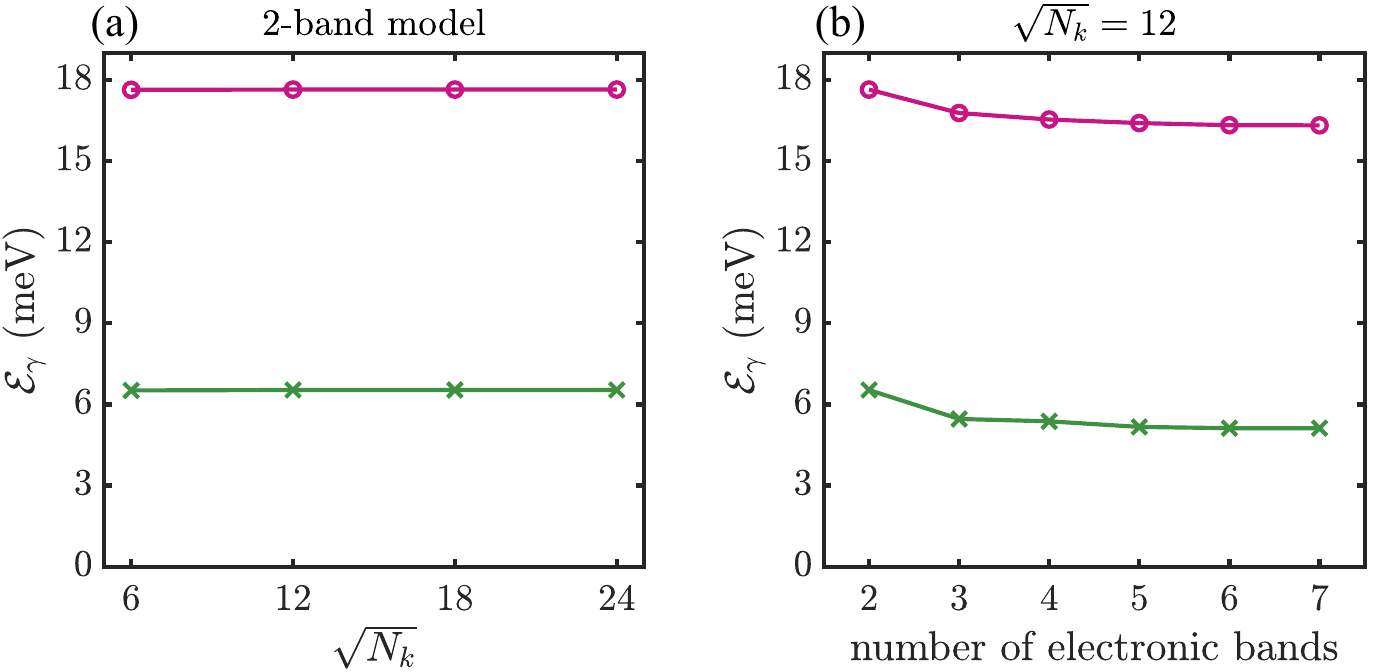}
\caption{Numerical results for the first two magnon energies at the $\gamma$ point, calculated using (a) different $\bm{k}$-mesh sizes with the number of electronic bands fixed at 2, and (b) different numbers of electronic bands with the $\bm{k}$-mesh size fixed at $N_k = 12 \times 12$.}
\label{figR2}
\end{figure}

We perform convergence tests for the magnon energies  versus the $\bm k$-mesh size $N_{k}$ and the number of electronic bands. The numerical convergence for the two lowest magnon energies at the $\gamma$ point is illustrated in Fig.~\ref{figR2}. We find that the magnon energies converge very fast with respect to $N_{k}$. While using only two electronic bands overestimates the magnon energies by about 1 meV, including approximately three bands is sufficient to achieve convergence. Our findings on numerical convergence are consistent with Ref.~\cite{Esquembre2025Magnon}, which also demonstrates that accurate magnon energies can be obtained using relatively few momentum points. Unless otherwise stated, we always present results using two electronic bands with $N_{k}=24$, as this choice provides a balance between numerical accuracy and computational efficiency.

\section{Calculation of magnon Chern number}\label{appb}

In the presence of electron-hole interactions, the
magnon states excited from $\tau=+$ to $\tau=-$ at $\nu=1$ can be parametrized as in Eq.~\eqref{exct}, where $f_{\bm k+\bm q,\lambda, -}^{\dagger}(f_{\bm k,1,+})$ can be expanded by field operator at real space $\bm r$ in layer $\ell$ as
\begin{equation}\label{fkrexp}
\begin{split}
&f_{\bm k+\bm q,\lambda,-}^{\dagger}=\sum_{\ell}\int d\bm r\psi^{\lambda}_{\bm k+\bm q,\ell,-}(\bm r) f_{\ell\bm r}^{\dagger},\\
&f_{\bm k,1,+}=\sum_{\ell}\int d\bm r[\psi^{1}_{\bm k,\ell,+}(\bm r)]^* f_{\ell\bm r}.
\end{split}
\end{equation}
Then
\begin{equation}\label{exctsm2}
|\chi, \bm q\rangle=\sum_{\ell,\ell'}\int d\bm r \int d\bm r' g_{\ell\ell'}^{\chi \bm q}(\bm r,\bm r') f_{\ell\bm r}^{\dagger} f_{\ell'\bm r'}|G\rangle,
\end{equation}
where $g_{\ell\ell'}^{\chi \bm q}(\bm r,\bm r')$ is the real-space
magnon Bloch wavefunction of Eq.~\eqref{rsmwf}. The periodic part of $g_{\ell\ell'}^{\chi \bm q}(\bm r,\bm r')$ can be expressed as
\begin{equation}
u_{\ell\ell'}^{\chi \bm q}(\bm r,\bm r')=e^{-i\bm q \frac{\bm r+\bm r'}{2}}\sum_{\bm k,\lambda}z^\lambda_{\bm k} (\chi, \bm q) \psi^{\lambda}_{\bm k+\bm q,\ell,-}(\bm r) [\psi^{1}_{\bm k,\ell',+}(\bm r')]^*.
\end{equation}
We compute the overlap integral as, 
\begin{equation}\label{uoverlap}
\begin{split}
\langle u^{\chi \bm q} | u^{\chi, \bm q+\bm q_0} \rangle
&=\sum_{\bm k \bm k'}\sum_{\lambda,\lambda'}
[z^{\lambda}_{\bm k}(\chi, \bm q)]^*
z^{\lambda'}_{\bm k'}(\chi, \bm q+\bm q_0)\\
&\int{d\bm{r}}e^{-i\frac{\bm q_0}{2} \bm r} \sum_{\ell} [\psi^{\lambda}_{\bm k+\bm q,\ell,-}(\bm r)]^* \psi^{\lambda'}_{\bm k'+\bm q+\bm q_0,\ell,-}(\bm r)\\ &\int{d\bm{r}'}e^{-i\frac{\bm q_0}{2} \bm r'} \sum_{\ell'} [\psi^{1}_{\bm k',\ell',+}(\bm r')]^* \psi^{1}_{\bm k,\ell',+}(\bm r')\\
&=\sum_{\bm k}\sum_{\lambda,\lambda'}
[z^{\lambda}_{\bm k}(\chi, \bm q)]^*
z^{\lambda'}_{\bm k-\frac{\bm q_0}{2}}(\chi, \bm q+\bm q_0)\\
&\langle u^{\lambda}_{\bm k+\bm q,-} | u^{\lambda'}_{\bm k+\bm q+\frac{\bm q_0}{2},-} \rangle
\langle u^{1}_{\bm k-\frac{\bm q_0}{2},+} | u^{1}_{\bm k,+} \rangle
\end{split}
\end{equation}
where $|u^{\lambda}_{\bm k,\tau} \rangle$  is the periodic part of $\psi_{\bm k, \tau}^{\lambda}$.
The Berry curvature of $\chi$-th magnon band at each $\bm q$ can be calculated using Eq.~\eqref{uoverlap} as follows,
\begin{equation}
\begin{split}
\Omega_{\chi\bm q}=&\frac{1}{\delta\mathcal{A}_0}\text{Arg} [\langle u^{\chi \bm q} | u^{\chi, \bm q+\bm q_1} \rangle
\langle u^{\chi, \bm q+ \bm q_1} | u^{\chi, \bm q+\bm q_1+\bm q_2} \rangle \\
&\langle u^{\chi, \bm q+\bm q_1+\bm q_2} | u^{\chi, \bm q+ \bm q_2} \rangle
\langle u^{\chi, \bm q+\bm q_2} | u^{\chi \bm q} \rangle ],  
\end{split}
\end{equation}
where $\bm q$, $\bm q+\bm q_1$, $\bm q+\bm q_1+\bm q_2$, and $\bm q+\bm q_2$ are four corners of a small plaquette with an area $\delta\mathcal{A}_0$ in the momentum mesh. The Chern number for the $\chi$-th magnon band can be expressed as
\begin{equation}\label{Chern_m}
\begin{split}
C_{\chi}=\frac{1}{2\pi}\int_{mBZ} d\bm q \ \Omega_{\chi\bm q}.
\end{split}
\end{equation}

\section{Derivation of effective spin model}\label{appc}

We present a microscopic derivation of the Lagrangian for the effective spin model in Eq.~\eqref{knbeph_main1}. To capture the main physics and make analytical progress, we consider a simplified Hamiltonian of Kane-Mele-Hubbard model, which only takes into account the onsite Hubbard interaction. This model can capture the essential physics of the QAHI state at $\nu=1$ and its collective excitation, as we show in the following.

The Kane-Mele-Hubbard model is given by,
\begin{equation}\label{KMH}
\begin{aligned}
&\hat{\mathcal{H}}_{\mathrm{KMH}} =\hat{\mathcal{H}}_{\mathrm{KM}}+\hat{\mathcal{H}}_U,\\
&\hat{\mathcal{H}}_{\mathrm{KM}}
=\sum_{\boldsymbol{k}} b_{\boldsymbol{k}}^{\dagger}\left[F_{\bm k}^0 \sigma_0 s_0+F_{\bm k}^x \sigma_x s_0+F_{\bm k}^y \sigma_y s_0+F_{\bm k}^z \sigma_z s_z\right] b_{\boldsymbol{k}}, \\
&\qquad\ =\sum_{\alpha\beta}\sum_{\bm k\tau}h^{\alpha\beta}_{\bm k\tau} b^\dagger_{\bm k\alpha\tau} b_{\bm k\beta\tau}, \\
&\hat{\mathcal{H}}_U =U \sum_{\alpha}\sum_{\bm R} \hat{n}_{\bm R\alpha \uparrow} \hat{n}_{\bm R\alpha\downarrow}
\end{aligned}
\end{equation}
where $\hat{\mathcal{H}}_{\mathrm{KM}}$ is obtained by Fourier transformation form Eq.~\eqref{thhtn}, and $b_{\boldsymbol{k}}=\{b_{\boldsymbol{k}A+},b_{\boldsymbol{k}A-},b_{\boldsymbol{k}B+},b_{\boldsymbol{k}B-}\}^{\mathrm{T}}$. Here $\sigma_{x,y,z}(\sigma_0)$ and $s_{x,y,z}(s_0)$ are, respectively, Pauli matrices (identity matrices) in the sublattice and spin spaces. $F^{0}_{\bm k}$ and $\bm F_{\bm k}=[F^{x}_{\bm k}\quad F^{y}_{\bm k}\quad F^{z}_{\bm k}]$ are real functions of the momentum $\boldsymbol{k}$. The  $F^{z}_{\bm k}$ term characterizes the Ising spin-orbit coupling in the Kane-Mele model. Due to the time-reversal symmetry,   $F_{\bm k}^0$ and $F_{\bm k}^x$ are even functions of $\bm k$, while $F_{\bm k}^y$ and $F_{\bm k}^z$ are odd functions of $\bm k$. We use $h_{\bm k\tau}$ to represent the $2\times 2$ single-particle Hamiltonian matrix for spin $\tau$. $\hat{\mathcal{H}}_U$ is the Hubbard term with onsite Coulomb repulsion $U$. 

At $\nu=1$, the Kane-Mele-Hubbard model can support the QAHI state with full spin (valley) polarization for a sufficiently large Hubbard $U$ \cite{Wu2019,InteractionDriven2023}. We assume the QAHI state is polarized to valley $\tau=+$ (spin up), and the corresponding density matrix can be expressed as,
\begin{equation}\label{expdensity}
\begin{aligned}
&\langle b^\dagger_{\bm kA +}b_{\bm kA+}\rangle=\frac{1}{2}(1-\frac{F^z_{\bm k}}{|\bm F_{\bm k}|}),\ 
\langle b^\dagger_{\bm kA+}b_{\bm kB+}\rangle=-\frac{F^x_{\bm k}+iF^y_{\bm k}}{2|\bm F_{\bm k}|},\\
&\langle b^\dagger_{\bm kB+}b_{\bm kA+}\rangle=-\frac{F^x_{\bm k}-iF^y_{\bm k}}{2|\bm F_{\bm k}|},\ 
\langle b^\dagger_{\bm kB+}b_{\bm kB+}\rangle=\frac{1}{2}(1+\frac{F^z_{\bm k}}{|\bm F_{\bm k}|}),\\
&\langle b^\dagger_{\bm k \alpha -}b_{\bm k \beta -}\rangle=\langle b^\dagger_{\bm k \alpha +}b_{\bm k \beta -}\rangle=\langle b^\dagger_{\bm k \alpha -}b_{\bm k \beta +}\rangle=0,
\end{aligned}
\end{equation}
where $|\bm F_{\bm k}|=\sqrt{(F_{\bm k}^x)^2+(F_{\bm k}^y)^2+(F_{\bm k}^z)^2}$.

We now construct an effective theory based on the Hamiltonian in Eq.~\eqref{KMH} and the following spin texture state~\cite{SpontaneousMoon1995}
\begin{equation}\label{spintexture}
\begin{split}
|\bm m\rangle&=e^{-i\mathcal{F}}|\uparrow\rangle
\approx (1-i\mathcal{F}-\frac{1}{2}\mathcal{F}^2)|\uparrow\rangle,\\
\mathcal{F}&=\sum_{\bm R\alpha}(m^{x}_{\bm R\alpha}S^{y}_{\bm R\alpha}-m^{y}_{\bm R\alpha}S^{x}_{\bm R\alpha})\\
&=\sum_{\bm q\alpha}(m^{\alpha,x}_{-\bm q}S^{\alpha,y}_{\bm q}-m^{\alpha,y}_{-\bm q}S^{\alpha,x}_{\bm q}),\\
&\bm S_{\bm R\alpha}=\frac{1}{\sqrt{N}}\sum_{\bm q}
e^{i\bm q \bm R}\bm S_{\bm q\alpha},\\
&\bm S_{\bm q\alpha}=\frac{1}{\sqrt{N}}\sum_{\bm k}\sum_{\tau_1,\tau_2}b^\dagger_{\bm k+\bm q,\alpha,\tau_1}\frac{\bm s_{\tau_1,\tau_2}}{2}b_{\bm k,\alpha,\tau_2},
\end{split}
\end{equation}
where $\bm m^{\alpha}_{\bm q}$ is the Fourier component of the unit vector $\bm m_{\bm R_\alpha}$ at sublattice $\alpha$ and unit cell $\bm R$, with small in-plane components $m^{x/y}_{\bm R\alpha}$, and $\bm S^{\alpha}_{\bm q}$ is the Fourier component of the local spin operator $\bm S_{\bm R\alpha}$. $\tau$ is the spin index.
The operator $e^{-i\mathcal{F}}$ rotates the local spin direction from $\hat{z}$ to $\bm m_{\bm R\alpha}$. The corresponding state $|\bm  m\rangle$ has a slowly varying spin texture relative to the  QAHI ground state $|\uparrow\rangle$ at $\nu=1$ with spin up polarization, where the average occupation number at every sublattice is $n_\alpha=\frac{1}{2}$.
By taking $m^{x/y}_{\bm R\alpha}$ as small parameters, 
the kinetic Berry phase $\mathcal{B}_{S}$ can be derived as
\begin{equation}\label{mptm3}
\begin{split}
\mathcal{B}_{S}=&\langle\bm m|i\hbar \partial_t|\bm m\rangle\\
=&-\frac{\hbar}{4}\sum_{\alpha\bm q}n_\alpha( m^{\alpha,x}_{\bm q}\partial_t{m}^{\alpha,y}_{-\bm q}-
m^{\alpha,y}_{\bm q}\partial_t{m}^{\alpha,x}_{-\bm q})\\
=&-\frac{\hbar}{4}\sum_{\alpha\bm R}n_\alpha( m^{x}_{\bm R\alpha}\partial_t{m}^{y}_{\bm R\alpha}-
m^{y}_{\bm R\alpha}\partial_t{m}^{x}_{\bm R\alpha}),
\end{split}
\end{equation}
which justifies Eq.~\eqref{kinetic_Berry}.

The energy of the spin texture state can be expanded in powers of $\mathcal{F}$ as, 
\begin{equation}\label{epsimm}
\begin{split}
\mathcal{E}[\bm m]=&\langle \bm m|\hat{\mathcal{H}}_{\mathrm{KMH}}|\bm m\rangle-\langle\hat{\mathcal{H}}_{\mathrm{KMH}}\rangle \\
\approx& i\langle[\mathcal{F}, \hat{\mathcal{H}}_{\mathrm{KMH}}]\rangle-\frac{1}{2}\langle[\mathcal{F},[\mathcal{F}, \hat{\mathcal{H}}_{\mathrm{KMH}}]]\rangle,
\end{split}
\end{equation}
where $\left\langle\cdots\right\rangle$ represents the ground state expectation value and the first term exactly vanishes. The second term can be divided into $\langle[\mathcal{F},[\mathcal{F}, \mathcal{H}_{\mathrm{KM}}]]\rangle$ and $\langle[\mathcal{F},[\mathcal{F}, \mathcal{H}_U]]\rangle$, where $\langle[\mathcal{F},[\mathcal{F}, \mathcal{H}_U]]\rangle=0$ because $\mathcal{H}_U$ preserves spin SU(2) symmetry at each site. 
The nonzero term can be expressed as 
\begin{equation}\label{FFHKM}
\begin{split}
\mathcal{E}[\bm m]=&-\frac{1}{2}\langle[\mathcal{F},[\mathcal{F},\hat{\mathcal{H}}_\mathrm{KM}]]\rangle\\
=&\sum_{\alpha}\sum_{\bm q}J^{\alpha\alpha}_{H}(\bm q) (m^{\alpha,x}_{\bm q} m^{\alpha,x}_{-\bm q}
+m^{\alpha,y}_{\bm q} m^{\alpha,y}_{-\bm q})\\
+&\sum_{\bm q}J^{AB}_{H}(\bm q) (m^{A,x}_{\bm q} m^{B,x}_{-\bm q}+
m^{A,y}_{\bm q} m^{B,y}_{-\bm q})\\
+&\sum_{\alpha}\sum_{\bm q}J^{\alpha\alpha}_{D}(\bm q)m^{\alpha,x}_{\bm q} m^{\alpha,y}_{-\bm q},
\end{split}
\end{equation}
where 
\begin{equation}\label{JJqconstant}
\begin{split}
J^{\alpha\alpha}_{H}(\bm q)=&\frac{1}{8N}\sum_{\bm k}
(h^{\alpha\alpha}_{\bm k+\bm q,-}+h^{\alpha\alpha}_{\bm k-\bm q,-})\langle b^\dagger_{\bm k\alpha+}
b_{\bm k\alpha+}\rangle\\
-&\frac{1}{4N}\sum_{\bm k\beta} \text{Re}[h^{\alpha\beta}_{\bm k+}\langle b^\dagger_{\bm k\alpha+}
b_{\bm k\beta+}\rangle],\\
J^{AB}_{H}(\bm q)=&\frac{1}{4N}\sum_{\bm k}h^{AB}_{\bm k+\bm q,-}\langle b^\dagger_{\bm k A+}b_{\bm k B+}\rangle \\
+&\frac{1}{4N}\sum_{\bm k}h^{BA}_{\bm k-\bm q,-}\langle b^\dagger_{\bm k B+}b_{\bm k A+}\rangle, \\
J^{\alpha\alpha}_{D}(\bm q)=&\frac{i}{4N}\sum_{\bm k}(h^{\alpha\alpha}_{\bm k+\bm q,-}-h^{\alpha\alpha}_{\bm k-\bm q,-})\langle b^\dagger_{\bm k\alpha+}b_{\bm k\alpha+}\rangle.
\end{split}
\end{equation}
It can be readily shown that
\begin{equation} \label{Jparity}
\begin{aligned}
&J_H^{\alpha \alpha}(-\bm q)=J_H^{\alpha \alpha}(\bm q) \in \text{Real},\\
&J_H^{AB}(-\bm q)=[J_H^{AB}(\bm q)]^*,\\
&J_D^{\alpha \alpha}(-\bm q)=[J_D^{\alpha \alpha}(\bm q)]^* \in \text{Imaginary}.
\end{aligned}
\end{equation}
Using the expression of $\langle b^\dagger_{\bm k\alpha+}b_{\bm k\beta+}\rangle$ in Eq.~\eqref{expdensity}, Eq.~\eqref{JJqconstant} can be further written as, 
\begin{equation}\label{Jqexpr}
\begin{aligned}
J^{AA}_{H}(\bm q)=&J^{BB}_{H}(\bm q)=\frac{1}{8N}\sum_{\bm k}[
|\bm F_{\bm k}|+(F^z_{\bm k+\bm q}+F^z_{\bm k-\bm q})\frac{F^z_{\bm k}}{2|\bm F_{\bm k}|}
],\\
J^{AB}_{H}(\bm q)
=&-\frac{1}{8N}\sum_{\bm k}(F^x_{\bm k+\bm q}-iF^y_{\bm k+\bm q})\frac{F^x_{\bm k}+iF^y_{\bm k}}{|\bm F_{\bm k}|}\\
-&\frac{1}{8N}\sum_{\bm k}(F^x_{\bm k-\bm q}+iF^y_{\bm k-\bm q})\frac{F^x_{\bm k}-iF^y_{\bm k}}{|\bm F_{\bm k}|},\\
J^{AA}_{D}(\bm q)=&-J^{BB}_{D}(\bm q)=\frac{i}{8N}\sum_{\bm k}(F^0_{\bm k-\bm q}-F^0_{\bm k+\bm q})\frac{F^z_{\bm k}}{|\bm F_{\bm k}|}.\\
\end{aligned}
\end{equation}
After Fourier transformation back to real space, we obtain
\begin{equation}\label{enecomp}
\begin{split}
\mathcal{E}[\bm m]=
&\sum_{\bm R\alpha}J^\alpha_0
[(m^x_{\bm R\alpha})^2+(m^y_{\bm R\alpha})^2]\\
+&\sum'_{\bm R \bm R'\alpha\beta} J^{\alpha\beta}_{\|}(\bm R'-\bm R)
(m^x_{\bm R\alpha}m^x_{\bm R'\beta}+m^y_{\bm R\alpha}m^y_{\bm R'\beta})\\
+&\sum'_{\bm R \bm R'\alpha}J^{\alpha\alpha}_{D}(\bm R'-\bm R)(\bm m_{\bm R\alpha}^x \bm m_{\bm R'\alpha}^y-\bm m_{\bm R\alpha}^y \bm m_{\bm R'\alpha}^x), 
\end{split}
\end{equation}
where the first line represents an onsite energy term, and the second and third lines are, respectively, Heisenberg coupling and DMI terms for the in-plane components $m^{x,y}$ at different sites, where the prime in the summation indicates that each bond is counted only once. 
Here the real space coupling constants are given by,
\begin{equation}\label{JRtq}
\begin{aligned}
&J^\alpha_0=\frac{1}{N}\sum_{\bm q}J^{\alpha\alpha}_H(\bm q)
=-\frac{1}{4}\sum_{\bm R\beta}^{\prime}\text{Re}[t^+_{\alpha\beta}(\bm R)\tilde{t}^+_{\alpha\beta}(\bm R)],\\
&J^{\alpha\alpha}_{\|}(\bm R)=\frac{2}{N}
\sum_{\bm q}e^{i\bm q\bm R} J^{\alpha\alpha}_{H}(\bm q)
=\frac{1}{2}\text{Re}[t^-_{\alpha\alpha}(\bm R)\tilde{t}^+_{\alpha\alpha}(\bm R)],\\
&J^{AB}_{\|}(\bm R)=\frac{1}{N}
\sum_{\bm q}e^{i\bm q\bm R} J^{AB}_{H}(\bm q)
=\frac{1}{2}\text{Re}[t^-_{AB}(\bm R)\tilde{t}^+_{AB}(\bm R)],\\
&J^{\alpha\alpha}_{D}(\bm R)=\frac{1}{N}
\sum_{\bm q}e^{i\bm q\bm R} J^{\alpha\alpha}_{D}(\bm q)
=-\frac{1}{2}\text{Im}[t^-_{\alpha\alpha}(\bm R)\tilde{t}^+_{\alpha\alpha}(\bm R)],\\
\end{aligned}
\end{equation}
where the prime in the summation of the first line denotes that the term with $\beta=\alpha,\ \bm R=\bm 0$ is excluded. Here $t^{\tau}_{\alpha\beta}(\bm R)$ represents the bare hopping in the Kane-Mele model, while $\tilde{t}^+_{\alpha\beta}(\bm R)=\langle b^\dagger_{\bm 0\alpha+} b_{\bm R\beta+} \rangle$ is the effective hopping measured with respect to the ground state. Note that $t^{\tau}_{\alpha\beta}(\bm R)$ has the unit of energy, while $\tilde{t}^+_{\alpha\beta}(\bm R)$ is dimensionless.

We now compare the energy functional in Eq.~\eqref{enecomp} with that in Eq.~\eqref{knbeph_main1}. Equation~\eqref{enecomp} can be viewed as Eq.~\eqref{knbeph_main1} expanded to the second order of the in-plane components $m^{x,y}$ under the approximation  of $m^z_{\bm R\alpha}\approx 1- \frac{1}{2}[(m^x_{\bm R\alpha})^2+(m^y_{\bm R\alpha})^2]$. Therefore, Eq.~\eqref{JRtq} presents a microscopic expression for the in-plane Heisenberg and DMI coupling constants. For the out-of-plane Heisenberg constants in Eq.~\eqref{knbeph_main1}, we have $-\frac{1}{2}\sum_{\bm R \beta} J_{z}^{\alpha \beta}(\bm R)=J_0^{\alpha}$. As shown in Eq.~\eqref{Jqexpr}, the DMI coupling constants are directly related to the Ising spin-orbit coupling term $F_{\bm k}^{z}$, which breaks the spin SU(2) symmetry down to U(1) symmetry. 

By combining Eqs.~\eqref{JzT0}-~\eqref{JD} and ~\eqref{JRtq}, we obtain the parameters for the magnon tight-binding model,
\begin{equation}
\begin{split}
\label{TRmagnon}
T_{\alpha \alpha}(\bm 0)=&8 J_0^{\alpha}=-2\sum_{\bm R\beta}^{\prime}\text{Re}[t^+_{\alpha\beta}(\bm R)\tilde{t}^+_{\alpha\beta}(\bm R)],\\
T_{\alpha\beta}(\bm R)=&\ 2\ t^-_{\alpha\beta}(\bm R)\tilde{t}^+_{\alpha\beta}(\bm R),
\end{split}
\end{equation}
where $T_{\alpha \alpha}(\bm 0)$ is the magnon onsite energy, and the second line is for hopping parameters between different sites. The magnon Wannier state is a bound state of a particle with spin down $(\tau=-)$ and a hole with spin up $(\tau=+)$. The magnon hopping parameter $T_{\alpha\beta}(\bm R)$ in Eq.~\eqref{TRmagnon} is proportional to the product of the bare particle hopping parameter $t^-_{\alpha\beta}(\bm R)$ and the dimensionless effective hole hopping parameter $\tilde{t}^+_{\alpha\beta}(\bm R)$, which provides a physical picture for the hopping of the bound state.

In summary, this derivation provides a microscopic justification of the Lagrangian for the lattice-based effective spin model in Eq.~\eqref{knbeph_main1}. In the presence of long-range of Coulomb interactions, the values of spin coupling constants are modified compared to those given by the expression in Eq.~\eqref{JRtq}, but the form of the Lagrangian are expected to be the same.

Throughout this work, we define magnons as collective excitations consisting of bound electron-hole pair (i.e., excitons) with spin flip—a definition that has been adopted in the literature for both moir\'e \cite{WuFengcheng2020Collective} and non-moir\'e systems \cite{Esquembre2025Magnon}. 
Excitons are known to obey bosonic statistics in the dilute limit, enabling phenomena such as exciton condensation. Similarly, magnons also follow bosonic statistics when their density is low, meaning that the total magnetization reversal relative to the ground state (i.e., the total number of magnons) remains small.
Moreover, the derived Lagrangian $\mathcal{L}_S=\mathcal{B}_S-\mathcal{E}[\bm m]$ takes the same structure as that describing magnon excitations in conventional spin systems, further establishing the bosonic nature of the magnons that we study.

\bibliography{main.bbl}

\end{document}